\documentclass[amsmath,amssymb,showkeys,superscriptaddress,aps,prd,10pt,onecolumn]{revtex4-2}
\usepackage{graphicx}
\usepackage[utf8]{inputenc}
\usepackage{multirow}
\usepackage{appendix}
\usepackage{float}
\usepackage{graphicx,epstopdf}
\usepackage{booktabs} 
\usepackage{xcolor}
\usepackage[colorlinks=true, urlcolor=blue, linkcolor=blue, citecolor=blue]{hyperref}
\DeclareUnicodeCharacter{039B}{\Lambda}

\begin{document}

\title{Magnetic moments of open bottom--charm molecular pentaquark octets}
\author{Halil Mutuk}%
\email[]{hmutuk@omu.edu.tr}
\affiliation{Department of Physics, Faculty of Sciences, Ondokuz Mayis University, 55200 Samsun, Türkiye}

\author{Xian-Wei Kang}%
\email[]{xwkang@bnu.edu.cn}
\affiliation{Key Laboratory of Beam Technology of the Ministry of Education,
School of Physics and Astronomy, Beijing Normal University, Beijing 100875, China}


\begin{abstract}
We present a comprehensive theoretical investigation of the magnetic moments of open heavy-flavor molecular pentaquarks with quark compositions $b\bar{c}qqq$ and $c\bar{b}qqq$ (where $q=u,d,s$). Employing a molecular picture in which the pentaquarks are treated as S-wave bound states of a heavy baryon and a meson, we systematically construct the complete spin--flavor wavefunctions for the two distinct SU(3)$_f$ octet representations, $8_{1f}$ and $8_{2f}$, arising from symmetric and antisymmetric light-diquark configurations, respectively. Within the framework of the constituent quark model, we calculate the magnetic moments of spin-parity configurations, $J^P = \frac{1}{2}^-(\frac{1}{2}^+\otimes 0^-)$ and $J^P = \frac{1}{2}^-, \frac{3}{2}^-(\frac{1}{2}^+\otimes 1^-)$, for each member of the $b\bar{c}$ and $c\bar{b}$ octets. Our results reveal a striking hierarchy: in the $8_{2f}$ representation, the $\frac{1}{2}^+\otimes 0^-$ states exhibit near-universal magnetic moments ($\mu \approx -0.062\,\mu_N$ for $b\bar{c}qqq$ and $\mu \approx +0.362\,\mu_N$ for $c\bar{b}qqq$), as a direct consequence of the spin-singlet light-diquark that suppresses light-quark contributions. In contrast, the $8_{1f}$ representation shows a broad spectrum of values with frequent sign changes, reflecting the active role of the symmetric light-diquark.  The clear differences between the $b\bar{c}$ and $c\bar{b}$ families demonstrate explicit heavy-quark flavor symmetry breaking in electromagnetic observables. These predictions provide a detailed set of electromagnetic benchmarks that can serve as discriminants for the internal flavor structure and spin configuration of future experimentally observed open heavy-flavor pentaquarks, offering valuable guidance for ongoing and future searches at facilities such as LHCb and Belle II.
\end{abstract}

\maketitle

\section{Introduction}\label{introduction}

The landscape of hadron spectroscopy has been profoundly reshaped by the discovery of exotic multiquark states, challenging the traditional meson and baryon configurations and offering new insights into the nonperturbative regime of Quantum Chromodynamics (QCD). The first landmark evidence for exotic hidden-charm pentaquark states was reported by the LHCb Collaboration in 2015 through the analysis of the \(J/\psi p\) invariant-mass distribution in the decay channel \(\Lambda_b^0 \to J/\psi K^- p\)~\cite{LHCb:2015yax}. Two resonant structures, \(P_c(4380)^+\) and \(P_c(4450)^+\), were identified. These findings constituted the first unambiguous observation of pentaquark candidates containing hidden-charm (\(c\bar{c}uud\)) and marked the beginning of a new era in the experimental exploration of multiquark hadrons.

Building upon this discovery, the LHCb Collaboration performed an updated amplitude analysis in 2019, revealing three distinct hidden-charm pentaquark states: \(P_c(4312)^+\), \(P_c(4440)^+\), and \(P_c(4457)^+\)~\cite{LHCb:2019kea}. The previously observed broad \(P_c(4450)^+\) resonance was resolved into two narrower states, \(P_c(4440)^+\) and \(P_c(4457)^+\). Following the convention for naming exotic hadrons~\cite{Gershon:2022xnn}, these three states are now denoted as
\begin{equation}
P_c(4312) \rightarrow P_{\psi}^N(4312), \quad
P_c(4440) \rightarrow P_{\psi}^N(4440), \quad
P_c(4457) \rightarrow P_{\psi}^N(4457).
\end{equation}

Further evidence for hidden-charm pentaquarks containing strangeness was reported by the LHCb Collaboration in 2021. A neutral resonance, \(P_{cs}(4459)^0\) (denoted as \(P_{\psi s}^{\Lambda}(4459)\) in the updated nomenclature), was found in the \(J/\psi \Lambda\) invariant-mass distribution of the decay \(\Xi_b^- \to J/\psi K^- \Lambda\), with a local significance of $3.1\sigma$ including systematic uncertainties~\cite{LHCb:2020jpq}. Independent evidence for the same state, with a local significance of $3.3\sigma$, has subsequently been reported by the Belle Collaboration in $\Upsilon(1S,2S)$ inclusive decays to $J/\psi\Lambda$, yielding a mass of $(4471.7\pm4.8\pm0.6)$\,MeV$/c^{2}$ and a width of $(21.9\pm13.1\pm2.7)$\,MeV~\cite{Belle:2025pey}. More recently, another strange hidden-charm pentaquark, \(P_{cs}(4338)^0\) (or \(P_{\psi s}^{\Lambda}(4338)\)), was observed by LHCb in the decay \(B^- \to J/\psi \Lambda \bar{p}\)~\cite{LHCb:2022ogu}.

Although the quantum numbers of these strange hidden-charm pentaquarks have not yet been experimentally determined, the preferred assignments suggested by amplitude analyses and by molecular interpretations are \(J^{P}=\tfrac{3}{2}^{-}\) for \(P_{\psi s}^{\Lambda}(4459)\) and \(J^{P}=\tfrac{1}{2}^{-}\) for \(P_{\psi s}^{\Lambda}(4338)\)~\cite{LHCb:2020jpq,LHCb:2022ogu,Chen:2016ryt,Wang:2019nvm,Yan:2022wuz}. These assignments do not follow from a generic spin--mass ordering: the two states are widely interpreted as molecules associated with \emph{different} meson--baryon channels, $\bar D^{*}\Xi_{c}$ for $P_{\psi s}^{\Lambda}(4459)$ and $\bar D\,\Xi_{c}$ for $P_{\psi s}^{\Lambda}(4338)$, with thresholds at $\approx 4477$ and $\approx 4336$\,MeV, respectively. The mass gap between the two candidates therefore reflects the gap between these underlying thresholds, not a hyperfine splitting; the same logic underlies our treatment of open bottom--charm pentaquarks below.

The discovery of hidden-charm pentaquarks has ignited a vibrant theoretical and experimental program. Significant progress has been made on magnetic moments in the hidden-charm sector~\cite{Wang:2016dzu,Ortiz-Pacheco:2018ccl,Xu:2020flp,Ozdem:2021ugy,Li:2021ryu,Ozdem:2023htj,Ozdem:2022kei,Gao:2021hmv,Guo:2023fih,Wang:2022nqs,Wang:2022tib,Ozdem:2024jty,Li:2024wxr,Li:2024jlq,Ozdem:2025fks,Ozdem:2025jda} and, to a lesser extent, in the hidden-bottom sector~\cite{Mutuk:2024jxf,Mutuk:2024ltc}. By contrast, pentaquarks containing both bottom (\(b\)) and charm (\(c\)) quarks --- with compositions $b\bar c qqq$ or $\bar c b qqq$ --- constitute a unique and relatively unexplored class of mixed heavy-flavor exotics that probes QCD dynamics in an intermediate mass regime. A complete color--magnetic SU(3)$_f$ classification of such heavy pentaquark multiplets was established in Ref.~\cite{Wu:2017weo}. Coupled-channel unitary analyses based on the local hidden-gauge approach have shown that meson--baryon interactions in the $\bar c b qqq$ and $b\bar c qqq$ systems dynamically generate multiple open-heavy resonances~\cite{Lin:2023iww}, and an extended local hidden-gauge study of $cc\bar q qs$, $bb\bar q qs$ and $bc\bar q qs$ systems has identified fourteen S-wave molecular candidates, including narrow bottom--charm states relevant to the present work~\cite{Wang:2025hhx}.

The magnetic moment is a fundamental electromagnetic property that encodes the distribution of charge and spin among a hadron's constituents. For pentaquarks, it depends sensitively on the relative alignment of heavy- and light-quark spins, the orbital configuration, and the overall color--flavor symmetry, and can therefore differ markedly across structural models even for states with similar masses and quantum numbers. This sensitivity makes the magnetic moment a powerful discriminant of the internal architecture of exotic hadrons, and is especially valuable in the bottom--charm case where the interplay of two distinct heavy quarks and three light quarks provides a richer structure than in single-heavy-flavor systems.

From an experimental standpoint, the search for bottom-charm pentaquarks is entering a promising era. The upgraded LHCb experiment and Belle~II are accumulating large statistics in heavy-flavor channels and can probe several plausible production mechanisms (fragmentation, photoproduction), with discovery channels such as $B_c$ plus light hadrons, $J/\psi \Lambda_b$, or $D^{(*)}\Sigma_b$. Although no bottom-charm pentaquark has been reported to date, theoretical guidance on their expected masses, widths, and distinguishing properties --- in particular their magnetic moments --- is essential to sharpen these searches and, in the event of a candidate, to elucidate its internal structure.

In this work, we focus specifically on bottom–anticharm (and charm–antibottom) pentaquarks within the molecular model, concentrating on the two distinct SU(3)$_f$ octet representations, $8_{1f}$ and $8_{2f}$, which originate from symmetric and antisymmetric light–diquark configurations in the baryon component. This framework provides a well-defined and computationally tractable approach for constructing pentaquark wave functions while respecting color confinement and Pauli statistics. In this picture, open bottom–charm pentaquarks are described as hadronic molecules composed of a singly heavy baryon and heavy--light meson, namely $(bqq)(\bar{c}q)$ for the $b\bar{c}$ sector and $(cqq)(\bar{b}q)$ for the $c\bar{b}$ sector. The interaction between these two hadrons is assumed to generate loosely bound states, predominantly in an S-wave configuration.

Throughout this work we denote the eight members of each open bottom--charm pentaquark octet by $P^{\,i}_{b\bar c}$ and $P^{\,i}_{c\bar b}$ ($i=1,\ldots,8$), following the compact labelling adopted in the \emph{Review of Particle Physics}~\cite{ParticleDataGroup:2024cfk} and in previous theoretical studies of $bc\bar q q q$ and $\bar b c q q q$ systems~\cite{Wu:2017weo,Lin:2023iww,Wang:2025hhx}. We emphasize that no exotic-hadron name has yet been assigned within the LHCb convention~\cite{Gershon:2022xnn} to any open bottom--charm pentaquark, simply because no such state has so far been experimentally observed. Should a candidate be reported in the future, the corresponding name in the LHCb scheme would take the form $P^{N}_{b\bar c}$ (for non-strange, isospin-doublet partners), $P^{\Sigma}_{b\bar c}$ or $P^{\Lambda}_{b\bar c}$ (for singly-strange members, depending on the light-quark isospin), and analogously for the $c\bar b$ family. The compact symbols $P^{\,i}_{b\bar c}$, $P^{\,i}_{c\bar b}$ used in the tables and figures of this paper are adopted purely for notational economy and do not preempt the LHCb naming convention.

The paper is organized as follows. Section~\ref{wavefunc} introduces the wave functions in the molecular picture, including the SU(3)$_f$
flavor wave functions and explicit spin wave functions. Section~\ref{expression} presents the magnetic moment formalism and its application to the pentaquark octets. Section~\ref{results} contains the numerical results and their physical interpretation. Section~\ref{final} summarizes our conclusions.

\section{Wave Functions}\label{wavefunc}

The electromagnetic structure of a hadron is determined by the internal configuration of its constituent quarks. For the open--flavor molecular
pentaquarks considered in this work, the total wave function factorizes into flavor, spin, color, and spatial components,
\begin{align}
\Psi
= \psi_{\mathrm{flavor}}
  \otimes \chi_{\mathrm{spin}}
  \otimes \xi_{\mathrm{color}}
  \otimes \eta_{\mathrm{space}} .
\end{align}
Fermi-Dirac statistics require the full wave function to be antisymmetric under exchange of identical light quarks. In the molecular picture, this
constraint is already satisfied within each constituent hadron. The baryon $(Qqq)$ contains two identical light quarks (in some flavor configurations),
and its color-flavor-spin wavefunction is appropriately antisymmetrized. The meson $(q\bar{Q}')$ contains no identical quarks. When these two
color-singlet hadrons bind to form a molecule, there is no additional antisymmetrization required across different hadrons, as the quarks are
confined to distinct spatial regions. The spatial part $\eta_{\mathrm{space}}$ for the ground-state S-wave molecule is symmetric, and the color
wavefunction $\xi_{\mathrm{color}}$ is simply the product of the two individual color singlets, which is overall symmetric.

A clarification regarding the role of the spatial component $\eta_{\mathrm{space}}$ in Eq.~(2) is in order. In the non-relativistic constituent quark model, the magnetic-moment operator $\hat{\bm{\mu}} = \sum_{i} (Q_{i}/2M_{i})\,\hat{\bm{\sigma}}_{i}$ acts only on the spin--flavor degrees of freedom of the individual quarks and carries no explicit spatial dependence. Consequently, the diagonal magnetic moment of a molecular state factorizes as
\begin{equation}
\langle\Psi|\hat{\mu}_{z}|\Psi\rangle
=\langle\eta_{\mathrm{space}}|\eta_{\mathrm{space}}\rangle\,
 \langle\chi_{\mathrm{spin}}\psi_{\mathrm{flavor}}\xi_{\mathrm{color}}|
        \hat{\mu}_{z}|
        \chi_{\mathrm{spin}}\psi_{\mathrm{flavor}}\xi_{\mathrm{color}}\rangle ,
\end{equation}
and since the spatial wave function is normalized, $\langle\eta_{\mathrm{space}}|\eta_{\mathrm{space}}\rangle = 1$, it contributes no overall factor to the result. The diagonal (static) magnetic moments computed in this work are therefore insensitive to the detailed shape of $\eta_{\mathrm{space}}$ and, in particular, are \emph{not} suppressed by the small inter-hadron spatial overlap that characterizes a loosely bound molecular configuration. Spatial-overlap effects do become important for \emph{off-diagonal} transition matrix elements $\langle\Psi'|\hat{\mu}_{z}|\Psi\rangle$ between molecules of different hadronic content, which are relevant for radiative-decay widths; we return to this point in Sec.~\ref{results}. The factorization adopted here is the standard prescription used throughout the constituent-quark-model literature on molecular pentaquarks~\cite{Li:2021ryu,Wang:2022tib,Wang:2022nqs,Li:2024wxr,Mutuk:2024jxf,Mutuk:2024ltc}.

In the molecular interpretation, open--bottom--charm pentaquarks arise from a singly heavy baryon and a heavy--light meson of the form
$(Qqq)(\bar Q' q)$, with $(Q,Q')=(b,c)$ or $(c,b)$. In principle, the same valence quark content $\bar Q' Q\,qqq$ admits a second hadronic decomposition, namely a light octet baryon combined with a heavy--heavy meson, $(qqq)(Q\bar Q')$, which for the open bottom--charm system corresponds to channels of the type $N B_{c}^{(*)}$, $\Lambda B_{c}^{(*)}$, $\Sigma B_{c}^{(*)}$, or $\Xi B_{c}^{(*)}$. We do not consider this configuration in the present work, for the following physical reason. Molecular binding in the meson--baryon sector is governed predominantly by light-meson exchange (one-pion, $\rho$, $\omega$, $\sigma$) and by light-quark rearrangement between the constituent hadrons. The $B_{c}^{(*)}$ meson, however, is an SU(3) flavor singlet containing no light quarks; consequently the $B_{c}B_{c}\pi$, $B_{c}B_{c}\rho$, $B_{c}B_{c}\omega$, and $B_{c}B_{c}\sigma$ vertices vanish at tree level, and the residual interaction with a light baryon is restricted to heavy-meson exchange and multi-pion or gluonic channels, all of which are strongly suppressed. As a result, the $(qqq)(Q\bar Q')$ configuration is not expected to generate dynamically bound molecular states, and is consistently excluded from the coupled-channel analyses of Refs.~\cite{Lin:2023iww,Wang:2025hhx} on which the present molecular assignment is based. We therefore focus exclusively on the $(Qqq)(\bar Q' q)$ class, which provides the only molecular configurations supported by the dynamical studies in the literature.

The flavor structure of the singly heavy baryon is governed by the SU(3) flavor symmetry of
the two light quarks. Symmetric light--diquark configurations,
\begin{align}
\{q_1 q_2\} = \frac{1}{\sqrt{2}}( q_1 q_2 + q_2 q_1 ),
\end{align}
belong to the sextet representation $6_f$, while antisymmetric diquarks,
\begin{align}
[q_1 q_2] = \frac{1}{\sqrt{2}}( q_1 q_2 - q_2 q_1 ),
\end{align}
belong to the antitriplet representation $\bar 3_f$. Combining the baryon with a heavy antimeson $(q\bar Q')$ in the $3_f$ representation generates
\begin{align}
6_f \otimes 3_f &= 10_f \oplus 8_{1f} , \\
\bar 3_f \otimes 3_f &= 8_{2f} \oplus 1_f ,
\end{align}
yielding the overall three--light--quark decomposition
\begin{align}
3\otimes3\otimes3 = 1 \oplus 8_1 \oplus 8_2 \oplus 10 .
\end{align}
Thus, two distinct SU(3)$_f$ flavor octets arise from Eqs.~(5)--(6): $8_{1f}$, built from symmetric light diquarks via $6_{f}\otimes 3_{f}$, and $8_{2f}$, built from antisymmetric light diquarks via $\bar 3_{f}\otimes 3_{f}$. In the present work we restrict our analysis to these two octet representations, for three concurrent reasons. First, all experimentally established hidden-charm pentaquark candidates --- $P_{\psi}^{N}(4312)$, $P_{\psi}^{N}(4440)$, $P_{\psi}^{N}(4457)$, $P_{\psi s}^{\Lambda}(4459)$, and $P_{\psi s}^{\Lambda}(4338)$ --- have been classified phenomenologically within SU(3)$_{f}$ octet multiplets~\cite{Wu:2017weo,Li:2024wxr}. Second, the coupled-channel dynamical analyses of Refs.~\cite{Lin:2023iww,Wang:2025hhx}, which provide the underlying motivation for treating open bottom--charm pentaquarks as hadronic molecules, identify the bulk of near-threshold $bc\bar q q q$ and $\bar bc\,qqq$ candidates in octet configurations. Third, restricting the calculation to the two octets allows a clean and physically transparent comparison between the symmetric and antisymmetric light-diquark sectors, which is one of the central themes of this paper. A complete treatment of the $10_{f}$ (built from $6_{f}\otimes 3_{f}$) and $1_{f}$ (built from $\bar 3_{f}\otimes 3_{f}$) representations is straightforward within the same formalism and is left for future investigation.

\begin{table*}[h]
\centering
\caption{Fully explicit spin--flavor wave functions of the $b\bar c$ molecular pentaquark octet.
The arrows denote quark spin projections. Here $\{q_1 q_2\}$ and $[q_1 q_2]$ denote symmetric and antisymmetric light--diquark
flavor combinations, respectively.}
\label{flavor_bcbar_spin}
\begin{tabular}{c c l}
\hline\hline
State & Representation & $\psi_{\rm flavor} \otimes \chi_{\rm spin}$ \\
\hline

$P_{b\bar c}^{1}$ & $8_{1f}$ &
$-\sqrt{\tfrac{1}{3}}
[(\{ud\}b)\otimes \tfrac{1}{\sqrt{6}}(\uparrow\downarrow\uparrow+\downarrow\uparrow\uparrow-2\uparrow\uparrow\downarrow)]
[(u\bar c)\otimes \tfrac{1}{\sqrt{2}}(\uparrow\downarrow-\downarrow\uparrow)]$
\\
&&
$+\sqrt{\tfrac{2}{3}}
[(\{uu\}b)\otimes \tfrac{1}{\sqrt{6}}(\uparrow\downarrow\uparrow+\downarrow\uparrow\uparrow-2\uparrow\uparrow\downarrow)]
[(d\bar c)\otimes \tfrac{1}{\sqrt{2}}(\uparrow\downarrow-\downarrow\uparrow)]$
\\[8pt]

$P_{b\bar c}^{1}$ & $8_{2f}$ &
$([ud]b\otimes \tfrac{1}{\sqrt{2}}(\uparrow\downarrow\uparrow-\downarrow\uparrow\uparrow))
(u\bar c\otimes \tfrac{1}{\sqrt{2}}(\uparrow\downarrow-\downarrow\uparrow))$
\\[10pt]

$P_{b\bar c}^{2}$ & $8_{1f}$ &
$\sqrt{\tfrac{1}{3}}
[(\{ud\}b)\otimes \tfrac{1}{\sqrt{6}}(\uparrow\downarrow\uparrow+\downarrow\uparrow\uparrow-2\uparrow\uparrow\downarrow)]
[(d\bar c)\otimes \tfrac{1}{\sqrt{2}}(\uparrow\downarrow-\downarrow\uparrow)]$
\\
&&
$-\sqrt{\tfrac{2}{3}}
[(\{dd\}b)\otimes \tfrac{1}{\sqrt{6}}(\uparrow\downarrow\uparrow+\downarrow\uparrow\uparrow-2\uparrow\uparrow\downarrow)]
[(u\bar c)\otimes \tfrac{1}{\sqrt{2}}(\uparrow\downarrow-\downarrow\uparrow)]$
\\[8pt]

$P_{b\bar c}^{2}$ & $8_{2f}$ &
$([ud]b\otimes \tfrac{1}{\sqrt{2}}(\uparrow\downarrow\uparrow-\downarrow\uparrow\uparrow))
(d\bar c\otimes \tfrac{1}{\sqrt{2}}(\uparrow\downarrow-\downarrow\uparrow))$
\\[10pt]

$P_{b\bar c}^{3}$ & $8_{1f}$ &
$\sqrt{\tfrac{1}{3}}
[(\{us\}b)\otimes \tfrac{1}{\sqrt{6}}(\uparrow\downarrow\uparrow+\downarrow\uparrow\uparrow-2\uparrow\uparrow\downarrow)]
[(u\bar c)\otimes \tfrac{1}{\sqrt{2}}(\uparrow\downarrow-\downarrow\uparrow)]$
\\
&&
$-\sqrt{\tfrac{2}{3}}
[(\{uu\}b)\otimes \tfrac{1}{\sqrt{6}}(\uparrow\downarrow\uparrow+\downarrow\uparrow\uparrow-2\uparrow\uparrow\downarrow)]
[(s\bar c)\otimes \tfrac{1}{\sqrt{2}}(\uparrow\downarrow-\downarrow\uparrow)]$
\\[8pt]

$P_{b\bar c}^{3}$ & $8_{2f}$ &
$([us]b\otimes \tfrac{1}{\sqrt{2}}(\uparrow\downarrow\uparrow-\downarrow\uparrow\uparrow))
(u\bar c\otimes \tfrac{1}{\sqrt{2}}(\uparrow\downarrow-\downarrow\uparrow))$
\\[10pt]

$P_{b\bar c}^{4}$ & $8_{1f}$ &
$\tfrac{1}{\sqrt{6}}
[(\{us\}b)\otimes \tfrac{1}{\sqrt{6}}(\uparrow\downarrow\uparrow+\downarrow\uparrow\uparrow-2\uparrow\uparrow\downarrow)]
[(d\bar c)\otimes \tfrac{1}{\sqrt{2}}(\uparrow\downarrow-\downarrow\uparrow)]$
\\
&&
$+\tfrac{1}{\sqrt{6}}
[(\{ds\}b)\otimes \tfrac{1}{\sqrt{6}}(\uparrow\downarrow\uparrow+\downarrow\uparrow\uparrow-2\uparrow\uparrow\downarrow)]
[(u\bar c)\otimes \tfrac{1}{\sqrt{2}}(\uparrow\downarrow-\downarrow\uparrow)]$
\\
&&
$-\sqrt{\tfrac{2}{3}}
[(\{ud\}b)\otimes \tfrac{1}{\sqrt{6}}(\uparrow\downarrow\uparrow+\downarrow\uparrow\uparrow-2\uparrow\uparrow\downarrow)]
[(s\bar c)\otimes \tfrac{1}{\sqrt{2}}(\uparrow\downarrow-\downarrow\uparrow)]$
\\[8pt]

$P_{b\bar c}^{4}$ & $8_{2f}$ &
$\tfrac{1}{\sqrt{2}}
([us]b\otimes \tfrac{1}{\sqrt{2}}(\uparrow\downarrow\uparrow-\downarrow\uparrow\uparrow))
(d\bar c\otimes \tfrac{1}{\sqrt{2}}(\uparrow\downarrow-\downarrow\uparrow))$
\\
&&
$+\tfrac{1}{\sqrt{2}}
([ds]b\otimes \tfrac{1}{\sqrt{2}}(\uparrow\downarrow\uparrow-\downarrow\uparrow\uparrow))
(u\bar c\otimes \tfrac{1}{\sqrt{2}}(\uparrow\downarrow-\downarrow\uparrow))$
\\[10pt]

$P_{b\bar c}^{5}$ & $8_{1f}$ &
$\tfrac{1}{\sqrt{2}}
[(\{us\}b)\otimes \tfrac{1}{\sqrt{6}}(\uparrow\downarrow\uparrow+\downarrow\uparrow\uparrow-2\uparrow\uparrow\downarrow)]
[(d\bar c)\otimes \tfrac{1}{\sqrt{2}}(\uparrow\downarrow-\downarrow\uparrow)]$
\\
&&
$-\tfrac{1}{\sqrt{2}}
[(\{ds\}b)\otimes \tfrac{1}{\sqrt{6}}(\uparrow\downarrow\uparrow+\downarrow\uparrow\uparrow-2\uparrow\uparrow\downarrow)]
[(u\bar c)\otimes \tfrac{1}{\sqrt{2}}(\uparrow\downarrow-\downarrow\uparrow)]$
\\[8pt]

$P_{b\bar c}^{5}$ & $8_{2f}$ &
$\tfrac{1}{\sqrt{6}}
([us]b\otimes \tfrac{1}{\sqrt{2}}(\uparrow\downarrow\uparrow-\downarrow\uparrow\uparrow))
(d\bar c\otimes \tfrac{1}{\sqrt{2}}(\uparrow\downarrow-\downarrow\uparrow))$
\\
&&
$-\tfrac{1}{\sqrt{6}}
([ds]b\otimes \tfrac{1}{\sqrt{2}}(\uparrow\downarrow\uparrow-\downarrow\uparrow\uparrow))
(u\bar c\otimes \tfrac{1}{\sqrt{2}}(\uparrow\downarrow-\downarrow\uparrow))$
\\
&&
$-\sqrt{\tfrac{2}{3}}
([ud]b\otimes \tfrac{1}{\sqrt{2}}(\uparrow\downarrow\uparrow-\downarrow\uparrow\uparrow))
(s\bar c\otimes \tfrac{1}{\sqrt{2}}(\uparrow\downarrow-\downarrow\uparrow))$
\\[10pt]

$P_{b\bar c}^{6}$ & $8_{1f}$ &
$\sqrt{\tfrac{1}{3}}
[(\{ds\}b)\otimes \tfrac{1}{\sqrt{6}}(\uparrow\downarrow\uparrow+\downarrow\uparrow\uparrow-2\uparrow\uparrow\downarrow)]
[(d\bar c)\otimes \tfrac{1}{\sqrt{2}}(\uparrow\downarrow-\downarrow\uparrow)]$
\\
&&
$-\sqrt{\tfrac{2}{3}}
[(\{dd\}b)\otimes \tfrac{1}{\sqrt{6}}(\uparrow\downarrow\uparrow+\downarrow\uparrow\uparrow-2\uparrow\uparrow\downarrow)]
[(s\bar c)\otimes \tfrac{1}{\sqrt{2}}(\uparrow\downarrow-\downarrow\uparrow)]$
\\[8pt]

$P_{b\bar c}^{6}$ & $8_{2f}$ &
$([ds]b\otimes \tfrac{1}{\sqrt{2}}(\uparrow\downarrow\uparrow-\downarrow\uparrow\uparrow))
(d\bar c\otimes \tfrac{1}{\sqrt{2}}(\uparrow\downarrow-\downarrow\uparrow))$
\\[10pt]

$P_{b\bar c}^{7}$ & $8_{1f}$ &
$\sqrt{\tfrac{1}{3}}
[(\{us\}b)\otimes \tfrac{1}{\sqrt{6}}(\uparrow\downarrow\uparrow+\downarrow\uparrow\uparrow-2\uparrow\uparrow\downarrow)]
[(s\bar c)\otimes \tfrac{1}{\sqrt{2}}(\uparrow\downarrow-\downarrow\uparrow)]$
\\
&&
$-\sqrt{\tfrac{2}{3}}
[(\{ss\}b)\otimes \tfrac{1}{\sqrt{6}}(\uparrow\downarrow\uparrow+\downarrow\uparrow\uparrow-2\uparrow\uparrow\downarrow)]
[(u\bar c)\otimes \tfrac{1}{\sqrt{2}}(\uparrow\downarrow-\downarrow\uparrow)]$
\\[8pt]

$P_{b\bar c}^{7}$ & $8_{2f}$ &
$([us]b\otimes \tfrac{1}{\sqrt{2}}(\uparrow\downarrow\uparrow-\downarrow\uparrow\uparrow))
(s\bar c\otimes \tfrac{1}{\sqrt{2}}(\uparrow\downarrow-\downarrow\uparrow))$
\\[10pt]

$P_{b\bar c}^{8}$ & $8_{1f}$ &
$\sqrt{\tfrac{1}{3}}
[(\{ds\}b)\otimes \tfrac{1}{\sqrt{6}}(\uparrow\downarrow\uparrow+\downarrow\uparrow\uparrow-2\uparrow\uparrow\downarrow)]
[(s\bar c)\otimes \tfrac{1}{\sqrt{2}}(\uparrow\downarrow-\downarrow\uparrow)]$
\\
&&
$-\sqrt{\tfrac{2}{3}}
[(\{ss\}b)\otimes \tfrac{1}{\sqrt{6}}(\uparrow\downarrow\uparrow+\downarrow\uparrow\uparrow-2\uparrow\uparrow\downarrow)]
[(d\bar c)\otimes \tfrac{1}{\sqrt{2}}(\uparrow\downarrow-\downarrow\uparrow)]$
\\[8pt]

$P_{b\bar c}^{8}$ & $8_{2f}$ &
$([ds]b\otimes \tfrac{1}{\sqrt{2}}(\uparrow\downarrow\uparrow-\downarrow\uparrow\uparrow))
(s\bar c\otimes \tfrac{1}{\sqrt{2}}(\uparrow\downarrow-\downarrow\uparrow))$

\\
\hline\hline
\end{tabular}
\end{table*}

\begin{table*}[h]
\centering
\caption{Fully explicit spin--flavor wave functions of the $c\bar b$ molecular pentaquark octet.
The arrows denote quark spin projections. Here $\{q_1 q_2\}$ and $[q_1 q_2]$ denote symmetric and antisymmetric light--diquark
flavor combinations, respectively.}
\label{flavor_cbbar_spin}
\begin{tabular}{c c l}
\hline\hline
State & Representation & $\psi_{\rm flavor} \otimes \chi_{\rm spin}$ \\
\hline

$P_{c\bar b}^{1}$ & $8_{1f}$ &
$-\sqrt{\tfrac{1}{3}}
[(\{ud\}c)\otimes \tfrac{1}{\sqrt{6}}(\uparrow\downarrow\uparrow+\downarrow\uparrow\uparrow-2\uparrow\uparrow\downarrow)]
[(u\bar b)\otimes \tfrac{1}{\sqrt{2}}(\uparrow\downarrow-\downarrow\uparrow)]$
\\
&&
$+\sqrt{\tfrac{2}{3}}
[(\{uu\}c)\otimes \tfrac{1}{\sqrt{6}}(\uparrow\downarrow\uparrow+\downarrow\uparrow\uparrow-2\uparrow\uparrow\downarrow)]
[(d\bar b)\otimes \tfrac{1}{\sqrt{2}}(\uparrow\downarrow-\downarrow\uparrow)]$
\\[8pt]

$P_{c\bar b}^{1}$ & $8_{2f}$ &
$([ud]c\otimes \tfrac{1}{\sqrt{2}}(\uparrow\downarrow\uparrow-\downarrow\uparrow\uparrow))
(u\bar b\otimes \tfrac{1}{\sqrt{2}}(\uparrow\downarrow-\downarrow\uparrow))$
\\[10pt]

$P_{c\bar b}^{2}$ & $8_{1f}$ &
$\sqrt{\tfrac{1}{3}}
[(\{ud\}c)\otimes \tfrac{1}{\sqrt{6}}(\uparrow\downarrow\uparrow+\downarrow\uparrow\uparrow-2\uparrow\uparrow\downarrow)]
[(d\bar b)\otimes \tfrac{1}{\sqrt{2}}(\uparrow\downarrow-\downarrow\uparrow)]$
\\
&&
$-\sqrt{\tfrac{2}{3}}
[(\{dd\}c)\otimes \tfrac{1}{\sqrt{6}}(\uparrow\downarrow\uparrow+\downarrow\uparrow\uparrow-2\uparrow\uparrow\downarrow)]
[(u\bar b)\otimes \tfrac{1}{\sqrt{2}}(\uparrow\downarrow-\downarrow\uparrow)]$
\\[8pt]

$P_{c\bar b}^{2}$ & $8_{2f}$ &
$([ud]c\otimes \tfrac{1}{\sqrt{2}}(\uparrow\downarrow\uparrow-\downarrow\uparrow\uparrow))
(d\bar b\otimes \tfrac{1}{\sqrt{2}}(\uparrow\downarrow-\downarrow\uparrow))$
\\[10pt]

$P_{c\bar b}^{3}$ & $8_{1f}$ &
$\sqrt{\tfrac{1}{3}}
[(\{us\}c)\otimes \tfrac{1}{\sqrt{6}}(\uparrow\downarrow\uparrow+\downarrow\uparrow\uparrow-2\uparrow\uparrow\downarrow)]
[(u\bar b)\otimes \tfrac{1}{\sqrt{2}}(\uparrow\downarrow-\downarrow\uparrow)]$
\\
&&
$-\sqrt{\tfrac{2}{3}}
[(\{uu\}c)\otimes \tfrac{1}{\sqrt{6}}(\uparrow\downarrow\uparrow+\downarrow\uparrow\uparrow-2\uparrow\uparrow\downarrow)]
[(s\bar b)\otimes \tfrac{1}{\sqrt{2}}(\uparrow\downarrow-\downarrow\uparrow)]$
\\[8pt]

$P_{c\bar b}^{3}$ & $8_{2f}$ &
$([us]c\otimes \tfrac{1}{\sqrt{2}}(\uparrow\downarrow\uparrow-\downarrow\uparrow\uparrow))
(u\bar b\otimes \tfrac{1}{\sqrt{2}}(\uparrow\downarrow-\downarrow\uparrow))$
\\[10pt]

$P_{c\bar b}^{4}$ & $8_{1f}$ &
$\tfrac{1}{\sqrt{6}}
[(\{us\}c)\otimes \tfrac{1}{\sqrt{6}}(\uparrow\downarrow\uparrow+\downarrow\uparrow\uparrow-2\uparrow\uparrow\downarrow)]
[(d\bar b)\otimes \tfrac{1}{\sqrt{2}}(\uparrow\downarrow-\downarrow\uparrow)]$
\\
&&
$+\tfrac{1}{\sqrt{6}}
[(\{ds\}c)\otimes \tfrac{1}{\sqrt{6}}(\uparrow\downarrow\uparrow+\downarrow\uparrow\uparrow-2\uparrow\uparrow\downarrow)]
[(u\bar b)\otimes \tfrac{1}{\sqrt{2}}(\uparrow\downarrow-\downarrow\uparrow)]$
\\
&&
$-\sqrt{\tfrac{2}{3}}
[(\{ud\}c)\otimes \tfrac{1}{\sqrt{6}}(\uparrow\downarrow\uparrow+\downarrow\uparrow\uparrow-2\uparrow\uparrow\downarrow)]
[(s\bar b)\otimes \tfrac{1}{\sqrt{2}}(\uparrow\downarrow-\downarrow\uparrow)]$
\\[8pt]

$P_{c\bar b}^{4}$ & $8_{2f}$ &
$\tfrac{1}{\sqrt{2}}
([us]c\otimes \tfrac{1}{\sqrt{2}}(\uparrow\downarrow\uparrow-\downarrow\uparrow\uparrow))
(d\bar b\otimes \tfrac{1}{\sqrt{2}}(\uparrow\downarrow-\downarrow\uparrow))$
\\
&&
$+\tfrac{1}{\sqrt{2}}
([ds]c\otimes \tfrac{1}{\sqrt{2}}(\uparrow\downarrow\uparrow-\downarrow\uparrow\uparrow))
(u\bar b\otimes \tfrac{1}{\sqrt{2}}(\uparrow\downarrow-\downarrow\uparrow))$
\\[10pt]

$P_{c\bar b}^{5}$ & $8_{1f}$ &
$\tfrac{1}{\sqrt{2}}
[(\{us\}c)\otimes \tfrac{1}{\sqrt{6}}(\uparrow\downarrow\uparrow+\downarrow\uparrow\uparrow-2\uparrow\uparrow\downarrow)]
[(d\bar b)\otimes \tfrac{1}{\sqrt{2}}(\uparrow\downarrow-\downarrow\uparrow)]$
\\
&&
$-\tfrac{1}{\sqrt{2}}
[(\{ds\}c)\otimes \tfrac{1}{\sqrt{6}}(\uparrow\downarrow\uparrow+\downarrow\uparrow\uparrow-2\uparrow\uparrow\downarrow)]
[(u\bar b)\otimes \tfrac{1}{\sqrt{2}}(\uparrow\downarrow-\downarrow\uparrow)]$
\\[8pt]

$P_{c\bar b}^{5}$ & $8_{2f}$ &
$\tfrac{1}{\sqrt{6}}
([us]c\otimes \tfrac{1}{\sqrt{2}}(\uparrow\downarrow\uparrow-\downarrow\uparrow\uparrow))
(d\bar b\otimes \tfrac{1}{\sqrt{2}}(\uparrow\downarrow-\downarrow\uparrow))$
\\
&&
$-\tfrac{1}{\sqrt{6}}
([ds]c\otimes \tfrac{1}{\sqrt{2}}(\uparrow\downarrow\uparrow-\downarrow\uparrow\uparrow))
(u\bar b\otimes \tfrac{1}{\sqrt{2}}(\uparrow\downarrow-\downarrow\uparrow))$
\\
&&
$-\sqrt{\tfrac{2}{3}}
([ud]c\otimes \tfrac{1}{\sqrt{2}}(\uparrow\downarrow\uparrow-\downarrow\uparrow\uparrow))
(s\bar b\otimes \tfrac{1}{\sqrt{2}}(\uparrow\downarrow-\downarrow\uparrow))$
\\[10pt]

$P_{c\bar b}^{6}$ & $8_{1f}$ &
$\sqrt{\tfrac{1}{3}}
[(\{ds\}c)\otimes \tfrac{1}{\sqrt{6}}(\uparrow\downarrow\uparrow+\downarrow\uparrow\uparrow-2\uparrow\uparrow\downarrow)]
[(d\bar b)\otimes \tfrac{1}{\sqrt{2}}(\uparrow\downarrow-\downarrow\uparrow)]$
\\
&&
$-\sqrt{\tfrac{2}{3}}
[(\{dd\}c)\otimes \tfrac{1}{\sqrt{6}}(\uparrow\downarrow\uparrow+\downarrow\uparrow\uparrow-2\uparrow\uparrow\downarrow)]
[(s\bar b)\otimes \tfrac{1}{\sqrt{2}}(\uparrow\downarrow-\downarrow\uparrow)]$
\\[8pt]

$P_{c\bar b}^{6}$ & $8_{2f}$ &
$([ds]c\otimes \tfrac{1}{\sqrt{2}}(\uparrow\downarrow\uparrow-\downarrow\uparrow\uparrow))
(d\bar b\otimes \tfrac{1}{\sqrt{2}}(\uparrow\downarrow-\downarrow\uparrow))$
\\[10pt]

$P_{c\bar b}^{7}$ & $8_{1f}$ &
$\sqrt{\tfrac{1}{3}}
[(\{us\}c)\otimes \tfrac{1}{\sqrt{6}}(\uparrow\downarrow\uparrow+\downarrow\uparrow\uparrow-2\uparrow\uparrow\downarrow)]
[(s\bar b)\otimes \tfrac{1}{\sqrt{2}}(\uparrow\downarrow-\downarrow\uparrow)]$
\\
&&
$-\sqrt{\tfrac{2}{3}}
[(\{ss\}c)\otimes \tfrac{1}{\sqrt{6}}(\uparrow\downarrow\uparrow+\downarrow\uparrow\uparrow-2\uparrow\uparrow\downarrow)]
[(u\bar b)\otimes \tfrac{1}{\sqrt{2}}(\uparrow\downarrow-\downarrow\uparrow)]$
\\[8pt]

$P_{c\bar b}^{7}$ & $8_{2f}$ &
$([us]c\otimes \tfrac{1}{\sqrt{2}}(\uparrow\downarrow\uparrow-\downarrow\uparrow\uparrow))
(s\bar b\otimes \tfrac{1}{\sqrt{2}}(\uparrow\downarrow-\downarrow\uparrow))$
\\[10pt]

$P_{c\bar b}^{8}$ & $8_{1f}$ &
$\sqrt{\tfrac{1}{3}}
[(\{ds\}c)\otimes \tfrac{1}{\sqrt{6}}(\uparrow\downarrow\uparrow+\downarrow\uparrow\uparrow-2\uparrow\uparrow\downarrow)]
[(s\bar b)\otimes \tfrac{1}{\sqrt{2}}(\uparrow\downarrow-\downarrow\uparrow)]$
\\
&&
$-\sqrt{\tfrac{2}{3}}
[(\{ss\}c)\otimes \tfrac{1}{\sqrt{6}}(\uparrow\downarrow\uparrow+\downarrow\uparrow\uparrow-2\uparrow\uparrow\downarrow)]
[(d\bar b)\otimes \tfrac{1}{\sqrt{2}}(\uparrow\downarrow-\downarrow\uparrow)]$
\\[8pt]

$P_{c\bar b}^{8}$ & $8_{2f}$ &
$([ds]c\otimes \tfrac{1}{\sqrt{2}}(\uparrow\downarrow\uparrow-\downarrow\uparrow\uparrow))
(s\bar b\otimes \tfrac{1}{\sqrt{2}}(\uparrow\downarrow-\downarrow\uparrow))$

\\
\hline\hline
\end{tabular}
\end{table*}

The spin structure follows from the quantum numbers of the constituents. Baryons built from a symmetric light-diquark configuration exist with
both $J_B^{P_B} = \frac{1}{2}^{+}$ and $\frac{3}{2}^{+}$, whereas those built from an antisymmetric light-diquark configuration are restricted to
$J_B^{P_B} = \frac{1}{2}^{+}$. The heavy-light mesons $(q\bar{Q}')$ appear with spin-parity $J_M^{P_M} = 0^{-}$ and $1^{-}$. For S--wave
molecules, the total spin--parity is obtained from
\begin{align}
J = S_B \oplus S_M ,
\qquad
P = P_B P_M ,
\end{align}
leading to the lowest configurations
\begin{align} \label{eq:JP}
J^P = \frac12^{-}(\tfrac12^{+}\otimes 0^{-}),\qquad
J^P = \frac12^{-}(\tfrac12^{+}\otimes 1^{-}),\qquad
J^P = \frac32^{-}(\tfrac12^{+}\otimes 1^{-}).
\end{align}

The SU(3)$_f$ spin--flavor wave functions of the $b\bar{c}$ and $c\bar{b}$ pentaquark octets are listed in Tables~\ref{flavor_bcbar_spin} and~\ref{flavor_cbbar_spin}, both at the hadronic level (baryon$\otimes$meson) and at the quark level (in terms of $\{q_1q_2\}$ and $[q_1q_2]$). The $8_{1f}$ representation comes from $6_f\otimes 3_f$ (symmetric diquarks) and the $8_{2f}$ from $\bar 3_f\otimes 3_f$ (antisymmetric diquarks); the orthogonality of the octet states follows from the underlying Clebsch--Gordan algebra.

The molecular assignment adopted here is supported by dynamical studies: coupled-channel calculations in the local hidden-gauge approach generate bound states near the relevant baryon--meson thresholds for the $\bar c b qqq$ and $b\bar c qqq$ systems~\cite{Lin:2023iww}, and the extended hidden-gauge analysis of Ref.~\cite{Wang:2025hhx} identifies fourteen S-wave molecular candidates --- including bottom--charm states --- close to the corresponding thresholds. The near-threshold character of these states is the standard criterion for a molecular interpretation, and the present work takes those results as motivation for computing their electromagnetic properties. Earlier predictions of heavy-baryon--heavy-meson molecules within the one-boson-exchange model are given in Ref.~\cite{Yang:2011wz}.

\section{Magnetic Moments}\label{expression}

We now compute the magnetic moments of the $b\bar c$ and $c\bar b$ molecular pentaquarks in the constituent quark model. The magnetic moment is in general a vector observable, $\hat{\bm\mu}$; following standard convention, the scalar quantity referred to as ``the magnetic moment'' $\mu$ of a hadron with total angular momentum $J$ is defined as
\begin{equation}
\mu \;\equiv\;\langle J,M_J=J|\,\hat\mu_z\,|J,M_J=J\rangle, \qquad \hat\mu_z\equiv\hat{\bm\mu}\cdot\hat e_z .
\end{equation}
Throughout this section all operator equations are written for the vector $\hat{\bm\mu}$; the numerical entries of Tables~\ref{tab:mag-bc} and~\ref{tab:mag-cb} correspond to the scalar projection $\mu$.

The hadronic magnetic moment receives intrinsic-spin and orbital contributions, $\hat{\bm\mu}=\hat{\bm\mu}_{\mathrm{spin}}+\hat{\bm\mu}_{\mathrm{orbital}}$. For the S-wave molecules considered here the relative orbital angular momentum vanishes ($L=0$), so only the spin contribution remains, with
\begin{equation}
\hat{\bm{\mu}}_{\mathrm{spin}}=\sum_{i}\frac{Q_i}{2M_i}\,\hat{\bm{\sigma}}_i ,
\end{equation}
where $Q_i$, $M_i$, $\hat{\bm{\sigma}}_i$ are the electric charge (in units of $e$), constituent mass, and Pauli spin vector of the $i$-th quark. Projecting onto the $z$-axis gives the elementary single-quark matrix elements $\langle q\!\uparrow|\hat\mu_z|q\!\uparrow\rangle = +Q_q/(2M_q)$ and $\langle q\!\downarrow|\hat\mu_z|q\!\downarrow\rangle = -Q_q/(2M_q)$.

In the molecular picture, an S-wave pentaquark is a coherent superposition of baryon--meson channels, and its magnetic-moment operator decomposes as
\begin{equation}
\hat{\bm{\mu}} = \hat{\bm{\mu}}_{B} + \hat{\bm{\mu}}_{M} ,
\end{equation}
where $\hat{\bm\mu}_B$ and $\hat{\bm\mu}_M$ are the (vector) operators of the baryon and meson components. This additive form follows directly from the factorization established below Eq.~(2): since $\hat{\bm\mu}$ is local at the quark level and the molecular spatial wave function is normalized, the diagonal pentaquark moment equals the sum of the constituent-hadron moments with no inter-hadron form-factor suppression.

To illustrate the quark-model calculation, consider the antisymmetric bottom baryon $\Lambda_b^{0}\equiv[ud]b$ with $J^{P}=\tfrac12^{+}$. Its spin--flavor wave function for $S_z=+\tfrac12$ is
\begin{equation}
|\Lambda_b^{0},\, S_z=+\tfrac12\rangle
= \frac{1}{\sqrt{2}}(ud - du)b \otimes \frac{1}{\sqrt{2}}
\bigl( \uparrow\downarrow\uparrow - \downarrow\uparrow\uparrow \bigr).
\end{equation}
Applying $\hat{\mu}_z=\sum_i(Q_i/2M_i)\sigma_z^{(i)}$ and evaluating the spin matrix elements gives the compact result $\mu_{\Lambda_b^{0}} = \mu_b$, where $\mu_q\equiv Q_q/(2M_q)$: the two light quarks in the antisymmetric $[ud]$ diquark couple to spin zero, so the moment is carried entirely by the heavy $b$ quark.

As a concrete illustration, consider the neutral open bottom--charm molecular pentaquark belonging to the $8_{1f}$ representation. Its
quark-level flavor wave function reads
\begin{equation}
P_{b\bar c}^{2}(8_{1f})
= \sqrt{\frac{1}{3}}\,\{ud\}b\,d\bar{c}
- \sqrt{\frac{2}{3}}\,\{dd\}b\,u\bar{c}.
\label{eq:P2_quarkflavor}
\end{equation}

For the S-wave configuration with $J^{P}=\tfrac12^{-}(\tfrac12^{+}\!\otimes\!0^{-})$, the meson carries zero spin and the total angular momentum is provided by the baryon; the explicit spin--flavor state with $M=+\tfrac12$ is given in Table~\ref{flavor_bcbar_spin}. Evaluating $\hat\mu_z=\sum_i\hat\mu_i$ on this state yields
\begin{align}
\mu\bigl[P_{b\bar c}^{2}(8_{1f})\bigr]
&= \tfrac{4}{3}\mu_d - \tfrac{1}{3}\mu_b
  + \tfrac{1}{3}\mu_d + \tfrac{2}{3}\mu_u + \mu_{\bar c},
\label{eq:P2_quark}
\end{align}
in agreement with the expression used in the numerical analysis; substituting the constituent-quark moments yields the prediction listed in Table~\ref{tab:mag-bc}.

\section{Results and Discussion}\label{results}

We use the constituent quark masses $m_u=m_d=361.8$~MeV, $m_s=540.4$~MeV, $m_c=1724.8$~MeV, and $m_b=5052.9$~MeV~\cite{Wu:2017weo}. The molecular masses of the open bottom--charm pentaquarks are not computed here; they are taken from Refs.~\cite{Lin:2023iww,Wang:2025hhx}, which predict the corresponding states to lie close to the relevant baryon--meson thresholds. For convenience, Tables~\ref{tab:mag-bc} and~\ref{tab:mag-cb} list, in addition to the magnetic moments, the dominant baryon--meson channels of each $P^{k}_{b\bar c}$ and $P^{k}_{c\bar b}$ and the corresponding thresholds from PDG averages~\cite{ParticleDataGroup:2024cfk}; the predicted molecular masses lie typically $10$--$50$\,MeV below these thresholds. Thus, for example, $P_{b\bar c}^{1}\,(8_{2f})$ should be sought in $\Lambda_{b}\bar D$ near $7484$\,MeV and $P_{c\bar b}^{1}\,(8_{2f})$ in $\Lambda_{c}B$ near $7566$\,MeV. The magnetic moments of the $b\bar c$ and $c\bar b$ octets exhibit a rich and systematic structure that links the electromagnetic properties directly to the underlying quark dynamics, spin coupling, and heavy-quark ordering, as discussed in the subsections below.

\subsection{Pseudoscalar channel: $\frac{1}{2}^+\otimes 0^-$, $J^P = \frac{1}{2}^-$}

The $\frac{1}{2}^+ \otimes 0^-$ configuration demonstrates a fundamental dichotomy between the two flavor representations. In the $8_{2f}$
representation, a near-perfect universality is observed. Every $b\bar c$ state from $P_{b\bar c}^1$ through $P_{b\bar c}^8$ possesses an identical
magnetic moment of $\mu = -0.062\,\mu_N$, and every $c\bar b$ state shares the value $\mu = +0.362\,\mu_N$. This constancy is a direct and powerful consequence of the antisymmetric nature of the light-diquark in the $8_{2f}$ wave function. In this configuration, the two light quarks are forced into a spin-singlet state, effectively silencing their contribution to the magnetic moment. Furthermore, the pseudoscalar meson
partner carries zero spin. Consequently, the magnetic moment of the entire molecular system is almost entirely dictated by the single heavy quark
(or antiquark) residing within the baryon component. The sign reversal between the $b\bar c$ and $c\bar b$ families encodes the change in the
dominant heavy constituent electric charge, while the magnitude difference reflects the inverse mass scaling: the charm quark larger magnetic
moment due to its lower mass dominates in the $c\bar b$ case. This provides an exceptionally clean electromagnetic signature for any state
belonging to the $8_{2f}$ molecular multiplet, making these predictions particularly robust and testable.

The behavior of the same $\frac{1}{2}^+ \otimes 0^-$ configurations in the $8_{1f}$ representation is markedly different. Here the light diquark
is in a symmetric spin-triplet configuration, allowing the light quarks to contribute actively to the total magnetic moment. As a result, the predicted values span a broad range including both positive and negative magnetic moments that depend sensitively on the flavor composition of each
state. For the $b\bar c$ sector, the values vary from large positive $\mu(P_{b\bar c}^1)=+1.749\,\mu_N$ and $\mu(P_{b\bar c}^3)=+1.812\,\mu_N$,
to moderate positive $\mu(P_{b\bar c}^4)=+0.372\,\mu_N$, and to negative values $\mu(P_{b\bar c}^2)=-0.555\,\mu_N$, $\mu(P_{b\bar c}^6)=-1.068\,\mu_N$, $\mu(P_{b\bar c}^7)=-0.238\,\mu_N$, and $\mu(P_{b\bar c}^8)=-0.814\,\mu_N$.
The state $P_{b\bar c}^5$ yields a comparatively small value $\mu(P_{b\bar c}^5)=-0.077\,\mu_N$, reflecting a partial cancellation
among contributions from different quark components.  A similar pattern appears in the $c\bar b$ sector: $\mu(P_{c\bar b}^1)=+1.607\,\mu_N$,
$\mu(P_{c\bar b}^3)=+1.671\,\mu_N$, $\mu(P_{c\bar b}^4)=+0.231\,\mu_N$, $\mu(P_{c\bar b}^2)=-0.697\,\mu_N$, $\mu(P_{c\bar b}^6)=-1.210\,\mu_N$, $\mu(P_{c\bar b}^7)=-0.380\,\mu_N$, $\mu(P_{c\bar b}^8)=-0.956\,\mu_N$, and $\mu(P_{c\bar b}^5)=-0.218\,\mu_N$. This wide spread of values originates from the interplay among the magnetic moments of the $u$, $d$, and $s$ quarks in the symmetric light-diquark configuration together with the heavy quark contributions. Since $\mu_q = Q_q/(2m_q)$, the different masses and charges of the $b$ and $c$ quarks directly produce quantitative differences between the $b\bar c$ and $c\bar b$ sectors, and the $8_{1f}$ representation consequently displays a much richer structure than the nearly universal pattern of the $8_{2f}$ case.

\subsection{Vector channel: $\frac{1}{2}^+\otimes 1^-$, $J^P = \frac{1}{2}^-$}

The introduction of a vector meson through the $\tfrac{1}{2}^+ \otimes 1^-$ configuration significantly enriches the observed patterns and highlights the critical role of spin coupling. The magnetic moments show substantial deviations from their pseudoscalar counterparts, often involving sign changes that reflect destructive interference between the baryon and vector-meson spin contributions.

A representative example is the state $P_{b\bar c}^1$. Its magnetic
moment changes from
$\mu(P_{b\bar c}^1)=+1.749\,\mu_N$
in the $\tfrac{1}{2}^+ \otimes 0^-$ channel to
$\mu(P_{b\bar c}^1)=-0.824\,\mu_N$
in the $\tfrac{1}{2}^+ \otimes 1^-$ channel within the $8_{1f}$
representation. In contrast, in the $8_{2f}$ representation the same
configuration yields
$\mu(P_{b\bar c}^1)=+0.931\,\mu_N$.
This demonstrates that the underlying flavor symmetry can qualitatively alter, and even reverse, the interference pattern that governs the magnetic moment. Such large shifts and sign reversals are a generic feature across both the $b\bar c$ and $c\bar b$ octets.

A comparison of corresponding states further illustrates the interplay between flavor structure and spin coupling. For the pair $P_{b\bar c}^2$ and $P_{c\bar b}^2$, one finds $\mu(P_{b\bar c}^2)=+0.519\,\mu_N$ and $\mu(P_{c\bar b}^2)=+0.273\,\mu_N$ for the $J=\tfrac{1}{2}$ states in the $8_{1f}$ representation, indicating similar constructive spin alignment. However, their $J=\tfrac{3}{2}$ moments differ noticeably: $\mu(P_{b\bar c}^2)=-0.054\,\mu_N$, while $\mu(P_{c\bar b}^2)=-0.635\,\mu_N$, showing that the heavy-quark content significantly affects higher-spin configurations.

The pair $P_{b\bar c}^3$ and $P_{c\bar b}^3$ exhibits a consistent pattern across both spin sectors. For $J=\tfrac{1}{2}$, the magnetic moments are negative, $\mu(P_{b\bar c}^3)=-0.719\,\mu_N$ and $\mu(P_{c\bar b}^3)=-0.965\,\mu_N$, while for $J=\tfrac{3}{2}$ they become positive,
$\mu(P_{b\bar c}^3)=+1.640\,\mu_N$ and $\mu(P_{c\bar b}^3)=+1.671\,\mu_N$. This behavior indicates a common spin-coupling mechanism in which the higher-spin configuration favors constructive alignment of the constituent magnetic moments despite the interchange of heavy quarks. The states $P_{b\bar c}^6$ and $P_{c\bar b}^6$ are distinguished by their large magnitudes across different configurations, reflecting a
flavor structure that enhances additive contributions. In particular, the $J=\tfrac{3}{2}$ state of $P_{b\bar c}^6$ reaches $\mu(P_{b\bar c}^6)=-2.104\,\mu_N$ in the $8_{1f}$ representation, one of the largest values in the spectrum.

\subsection{Vector channel: $\frac{1}{2}^+\otimes 1^-$, $J^P = \frac{3}{2}^-$}

The $J^P=\tfrac{3}{2}^-$ states, originating from maximal spin alignment in the $\tfrac{1}{2}^+ \otimes 1^-$ coupling, systematically exhibit the
largest magnetic moments in magnitude across both SU(3)$_f$ representations and both heavy-quark sectors. This enhancement follows
from the coherent addition of spin contributions from the baryon and the vector meson.

The effect is particularly pronounced in the $c\bar b$ sector, where the charm quark carries a comparatively larger intrinsic magnetic moment.
A representative example is the state $P_{c\bar b}^1$, for which
\begin{align}
\mu(P_{c\bar b}^1)=+2.533\,\mu_N
\end{align}
in the $8_{1f}$ representation for $J=\tfrac{3}{2}$, making it one of the largest magnetic moments in the spectrum.

This behavior is generic, indicating that the $J=\tfrac{3}{2}$ states are the most electromagnetically enhanced configurations and therefore the
most favorable candidates for observation in electromagnetic processes. The dependence on the flavor representation remains significant. For
instance, the state $P_{c\bar b}^6$ shows a pronounced variation between
the two octets:
\begin{align}
\mu(P_{c\bar b}^6) &= -0.957\,\mu_N \quad (8_{1f}), \\
\mu(P_{c\bar b}^6) &= +2.153\,\mu_N \quad (8_{2f}).
\end{align}
The large magnitude difference and the sign reversal clearly demonstrate the sensitivity of magnetic moments to the underlying flavor structure.
Such behavior provides a powerful diagnostic tool for distinguishing between the $8_{1f}$ and $8_{2f}$ assignments of experimentally observed
states.

\subsection{Theoretical implications}

Several implications follow from these results. First, the systematic gap between the $8_{1f}$ and $8_{2f}$ predictions establishes the magnetic moment as a useful discriminator of the internal flavor structure of molecular pentaquarks. Second, the differences between the $b\bar c$ and $c\bar b$ families across all spin--parity configurations constitute explicit evidence of heavy-quark flavor symmetry breaking in electromagnetic observables: since $\mu\propto Q/m$, interchanging the roles of the $b$ and $c$ quarks between the baryon and meson modifies the relative quark-sector contributions in a way not captured by simple symmetry arguments. The sign change of the universal $8_{2f}$ pseudoscalar moment between families --- from $\mu\approx-0.062\,\mu_N$ for $b\bar c$ to $\mu\approx+0.362\,\mu_N$ for $c\bar b$ --- is the cleanest manifestation of this breaking. Third, the sizable splitting between $J=1/2$ and $J=3/2$ moments arising from the same hadronic configuration provides a potential indicator for spin determination, complementary to angular-distribution analyses~\cite{Jacob:1959at,LHCb:2024eyx,Kang:2009sb,Kang:2010td,Charles:2009ig}. Fourth, the near-vanishing $\mu(P_{b\bar c}^5)=-0.077\,\mu_N$ in the $8_{1f}$ pseudoscalar channel reflects a fine balance between light- and heavy-quark contributions within the adopted mass scheme.

Direct measurements of static moments for short-lived pentaquarks are currently beyond reach, but indirect access is possible through radiative decay widths, proportional to the cube of the photon energy and the square of the transition magnetic moment~\cite{Wang:2019mhm}. The diagonal moments tabulated here are insufficient for that purpose: radiative widths require off-diagonal matrix elements $\langle\Psi'|\hat\mu_z|\Psi\rangle$ between states of different spin--parity configuration, which explicitly involve the spatial overlap $\langle\eta'_{\mathrm{space}}|\eta_{\mathrm{space}}\rangle$ between the initial and final molecules. In the loosely-bound limit this overlap acts as a form-factor-like suppression sensitive to the binding dynamics; its quantitative treatment, requiring an explicit spatial wave function, is left for future work. Production asymmetries in polarized collisions at LHCb, Belle~II and the Electron--Ion Collider may provide further indirect probes, and the predictions tabulated here serve as benchmarks for such future analyses.

\begin{table*}[h]
\caption{ \label{tab:mag-bc}Magnetic moments of the $b\bar c$ octet molecular pentaquark family states, in units of the nuclear magneton $\mu_N$. The third column lists the dominant baryon--meson channels for each representation (full spin--flavor combinations are given in Table~\ref{flavor_bcbar_spin}). The fourth column gives the approximate thresholds (sums of constituent masses from PDG averages~\cite{ParticleDataGroup:2024cfk}) for the pseudoscalar (PS) and vector (V) meson channels.}
\begin{ruledtabular}
\begin{tabular}{ccccccc}
State & $J_B^{P_b}\!\otimes\!J_M^{P_m}$ & Channels [$8_{2f}$\,/\,$8_{1f}$] & Threshold (MeV)\ PS\,/\,V & $I(J^P)$ & $\mu(8_{1f})$ & $\mu(8_{2f})$ \\
\hline
$P_{b\bar c}^1$& $\tfrac12^+\!\otimes\!0^-$ & $\Lambda_b\bar D$\,/\,$\Sigma_b\bar D$  & $7484\,/\,7676$ &  $\tfrac12(\tfrac12^-)$ & $1.749$  & $-0.062$ \\
              & $\tfrac12^+\!\otimes\!1^-$ &                                          & $7627\,/\,7820$ &  $\tfrac12(\tfrac12^-)$ & $-0.824$ & $0.931$  \\
              &                            &                                          &                 &  $\tfrac12(\tfrac32^-)$ & $1.386$  & $1.304$  \\ \hline
$P_{b\bar c}^2$& $\tfrac12^+\!\otimes\!0^-$ & $\Lambda_b D$\,/\,$\Sigma_b D$           & $7489\,/\,7681$ &  $\tfrac12(\tfrac12^-)$ & $-0.555$ & $-0.062$ \\
              & $\tfrac12^+\!\otimes\!1^-$ &                                          & $7630\,/\,7822$ &  $\tfrac12(\tfrac12^-)$ & $0.519$  & $-0.797$ \\
              &                            &                                          &                 &  $\tfrac12(\tfrac32^-)$ & $-0.054$ & $-1.289$ \\ \hline
$P_{b\bar c}^3$& $\tfrac12^+\!\otimes\!0^-$ & $\Xi_b\bar D$\,/\,$\Xi_b'\bar D,\Sigma_b D_s$ & $7657\,/\,7779$ &  $1(\tfrac12^-)$       & $1.812$  & $-0.062$ \\
              & $\tfrac12^+\!\otimes\!1^-$ &                                          & $7799\,/\,7923$ &  $1(\tfrac12^-)$       & $-0.719$ & $0.931$  \\
              &                            &                                          &                 &  $1(\tfrac32^-)$       & $1.640$  & $1.304$  \\ \hline
$P_{b\bar c}^4$& $\tfrac12^+\!\otimes\!0^-$ & $\Xi_b D,\Xi_b\bar D$\,/\,$\Xi_b'\bar D,\Sigma_b D_s$  & $7662\,/\,7784$ &  $1(\tfrac12^-)$       & $0.372$  & $-0.062$ \\
              & $\tfrac12^+\!\otimes\!1^-$ &                                          & $7803\,/\,7925$ &  $1(\tfrac12^-)$       & $-0.527$ & $0.670$  \\
              &                            &                                          &                 &  $1(\tfrac32^-)$       & $-0.232$ & $0.008$  \\ \hline
$P_{b\bar c}^5$& $\tfrac12^+\!\otimes\!0^-$ & $\Lambda_b D_s,\Xi_b D$\,/\,$\Xi_b'\bar D,\Sigma_b D_s$ & $7588\,/\,7779$ &  $0(\tfrac12^-)$       & $-0.077$ & $-0.062$ \\
              & $\tfrac12^+\!\otimes\!1^-$ &                                          & $7731\,/\,7923$ &  $0(\tfrac12^-)$       & $0.072$  & $-0.382$ \\
              &                            &                                          &                 &  $0(\tfrac32^-)$       & $-0.007$ & $-0.666$ \\ \hline
$P_{b\bar c}^6$& $\tfrac12^+\!\otimes\!0^-$ & $\Xi_b D$\,/\,$\Xi_b' D,\Sigma_b D_s$    & $7667\,/\,7784$ &  $1(\tfrac12^-)$       & $-1.068$ & $-0.062$ \\
              & $\tfrac12^+\!\otimes\!1^-$ &                                          & $7807\,/\,7925$ &  $1(\tfrac12^-)$       & $-0.335$ & $-0.797$ \\
              &                            &                                          &                 &  $1(\tfrac32^-)$       & $-2.104$ & $-1.289$ \\ \hline
$P_{b\bar c}^7$& $\tfrac12^+\!\otimes\!0^-$ & $\Xi_b D_s$\,/\,$\Xi_b' D_s,\Omega_b\bar D$  & $7760\,/\,7903$ &  $\tfrac12(\tfrac12^-)$ & $-0.238$ & $-0.062$ \\
              & $\tfrac12^+\!\otimes\!1^-$ &                                          & $7904\,/\,8049$ &  $\tfrac12(\tfrac12^-)$ & $0.477$  & $-0.607$ \\
              &                            &                                          &                 &  $\tfrac12(\tfrac32^-)$ & $0.359$  & $-1.003$ \\ \hline
$P_{b\bar c}^8$& $\tfrac12^+\!\otimes\!0^-$ & $\Xi_b D_s$\,/\,$\Xi_b' D_s,\Omega_b D$ & $7765\,/\,7903$ &  $\tfrac12(\tfrac12^-)$ & $-0.814$ & $-0.062$ \\
              & $\tfrac12^+\!\otimes\!1^-$ &                                          & $7907\,/\,8049$ &  $\tfrac12(\tfrac12^-)$ & $-0.483$ & $-0.607$ \\
              &                            &                                          &                 &  $\tfrac12(\tfrac32^-)$ & $-1.946$ & $-1.003$ \\
\end{tabular}
\end{ruledtabular}
\end{table*}

\begin{table*}[h]
\caption{ \label{tab:mag-cb}Magnetic moments of the $c\bar b$ octet molecular pentaquark family states, in units of the nuclear magneton $\mu_N$. The third column lists the dominant baryon--meson channels for each representation (full spin--flavor combinations are given in Table~\ref{flavor_cbbar_spin}). The fourth column gives the approximate thresholds (sums of constituent masses from PDG averages~\cite{ParticleDataGroup:2024cfk}) for the pseudoscalar (PS) and vector (V) meson channels.}
\begin{ruledtabular}
\begin{tabular}{ccccccc}
State & $J_B^{P_b}\!\otimes\!J_M^{P_m}$ & Channels [$8_{2f}$\,/\,$8_{1f}$] & Threshold (MeV)\ PS\,/\,V & $I(J^P)$ & $\mu(8_{1f})$ & $\mu(8_{2f})$ \\
\hline
$P_{c\bar b}^1$& $\tfrac12^+\!\otimes\!0^-$ & $\Lambda_c B$\,/\,$\Sigma_c B$  & $7566\,/\,7732$ &  $\tfrac12(\tfrac12^-)$ & $1.607$  & $0.362$  \\
              & $\tfrac12^+\!\otimes\!1^-$ &                                  & $7611\,/\,7777$ &  $\tfrac12(\tfrac12^-)$ & $0.081$  & $-0.656$ \\
              &                            &                                  &                 &  $\tfrac12(\tfrac32^-)$ & $2.533$  & $-0.440$ \\ \hline
$P_{c\bar b}^2$& $\tfrac12^+\!\otimes\!0^-$ & $\Lambda_c B$\,/\,$\Sigma_c B$  & $7566\,/\,7732$ &  $\tfrac12(\tfrac12^-)$ & $-0.697$ & $0.362$  \\
              & $\tfrac12^+\!\otimes\!1^-$ &                                  & $7611\,/\,7777$ &  $\tfrac12(\tfrac12^-)$ & $0.273$  & $1.073$  \\
              &                            &                                  &                 &  $\tfrac12(\tfrac32^-)$ & $-0.635$ & $2.153$  \\ \hline
$P_{c\bar b}^3$& $\tfrac12^+\!\otimes\!0^-$ & $\Xi_c B$\,/\,$\Xi_c' B,\Sigma_c B_s$ & $7747\,/\,7857$ &  $1(\tfrac12^-)$       & $1.671$  & $0.362$  \\
              & $\tfrac12^+\!\otimes\!1^-$ &                                  & $7792\,/\,7903$ &  $1(\tfrac12^-)$       & $-0.965$ & $-0.656$ \\
              &                            &                                  &                 &  $1(\tfrac32^-)$       & $1.060$  & $-0.440$ \\ \hline
$P_{c\bar b}^4$& $\tfrac12^+\!\otimes\!0^-$ & $\Xi_c B$\,/\,$\Xi_c' B,\Sigma_c B_s$ & $7747\,/\,7857$ &  $1(\tfrac12^-)$       & $0.231$  & $0.362$  \\
              & $\tfrac12^+\!\otimes\!1^-$ &                                  & $7792\,/\,7903$ &  $1(\tfrac12^-)$       & $-0.197$ & $0.208$  \\
              &                            &                                  &                 &  $1(\tfrac32^-)$       & $0.051$  & $0.856$  \\ \hline
$P_{c\bar b}^5$& $\tfrac12^+\!\otimes\!0^-$ & $\Lambda_c B_s,\Xi_c B$\,/\,$\Xi_c' B,\Sigma_c B_s$  & $7653\,/\,7857$ &  $0(\tfrac12^-)$       & $-0.218$ & $0.362$  \\
              & $\tfrac12^+\!\otimes\!1^-$ &                                  & $7702\,/\,7903$ &  $0(\tfrac12^-)$       & $0.402$  & $-0.241$ \\
              &                            &                                  &                 &  $0(\tfrac32^-)$       & $0.275$  & $0.183$  \\ \hline
$P_{c\bar b}^6$& $\tfrac12^+\!\otimes\!0^-$ & $\Xi_c B$\,/\,$\Xi_c' B,\Sigma_c B_s$ & $7750\,/\,7858$ &  $1(\tfrac12^-)$       & $-1.210$ & $0.362$  \\
              & $\tfrac12^+\!\otimes\!1^-$ &                                  & $7795\,/\,7903$ &  $1(\tfrac12^-)$       & $0.571$  & $1.073$  \\
              &                            &                                  &                 &  $1(\tfrac32^-)$       & $-0.957$ & $2.153$  \\ \hline
$P_{c\bar b}^7$& $\tfrac12^+\!\otimes\!0^-$ & $\Xi_c B_s$\,/\,$\Xi_c' B_s,\Omega_c B$  & $7834\,/\,7945$ &  $\tfrac12(\tfrac12^-)$ & $-0.380$ & $0.362$  \\
              & $\tfrac12^+\!\otimes\!1^-$ &                                  & $7883\,/\,7994$ &  $\tfrac12(\tfrac12^-)$ & $-0.345$ & $-0.465$ \\
              &                            &                                  &                 &  $\tfrac12(\tfrac32^-)$ & $-1.087$ & $-0.154$ \\ \hline
$P_{c\bar b}^8$& $\tfrac12^+\!\otimes\!0^-$ & $\Xi_c B_s$\,/\,$\Xi_c' B_s,\Omega_c B$  & $7837\,/\,7945$ &  $\tfrac12(\tfrac12^-)$ & $-0.956$ & $0.362$  \\
              & $\tfrac12^+\!\otimes\!1^-$ &                                  & $7886\,/\,7994$ &  $\tfrac12(\tfrac12^-)$ & $0.999$  & $-0.465$ \\
              &                            &                                  &                 &  $\tfrac12(\tfrac32^-)$ & $0.065$  & $-0.154$ \\
\end{tabular}
\end{ruledtabular}
\end{table*}

\begin{figure}[htb!]
\includegraphics[width=0.45\textwidth]{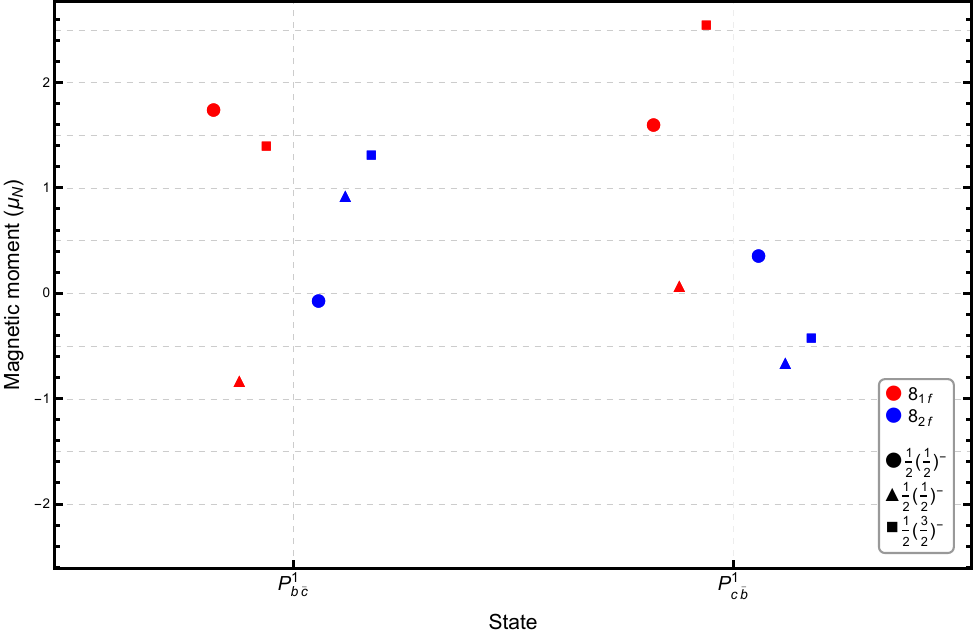}
\includegraphics[width=0.45\textwidth]{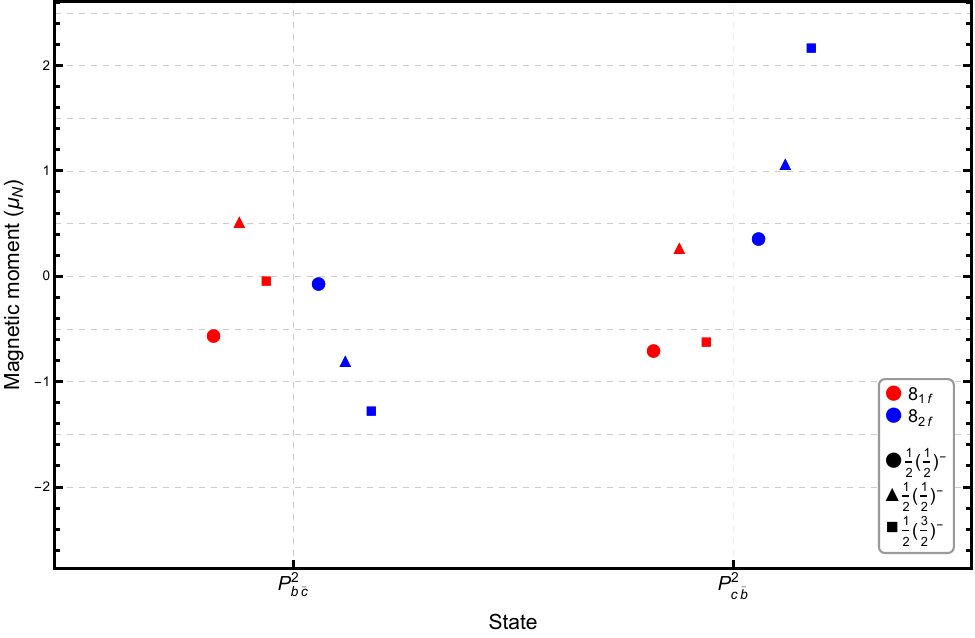}\\
\includegraphics[width=0.45\textwidth]{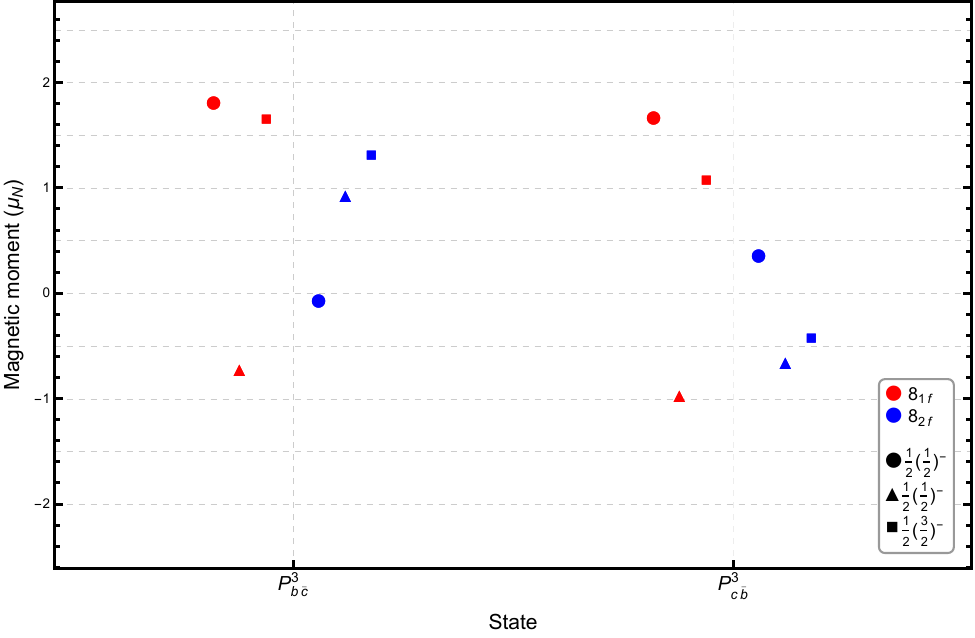}
\includegraphics[width=0.45\textwidth]{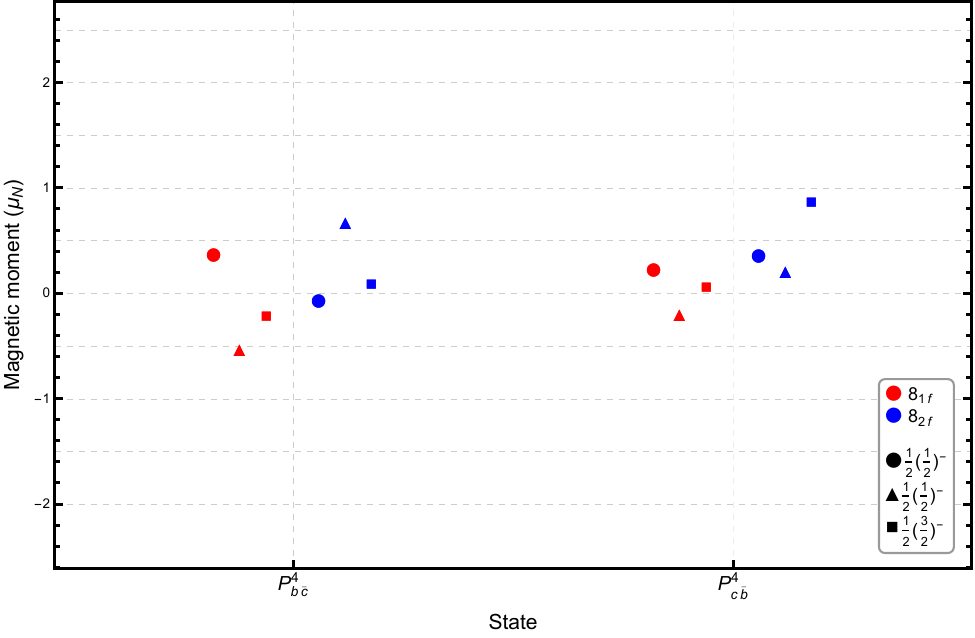}\\
\includegraphics[width=0.45\textwidth]{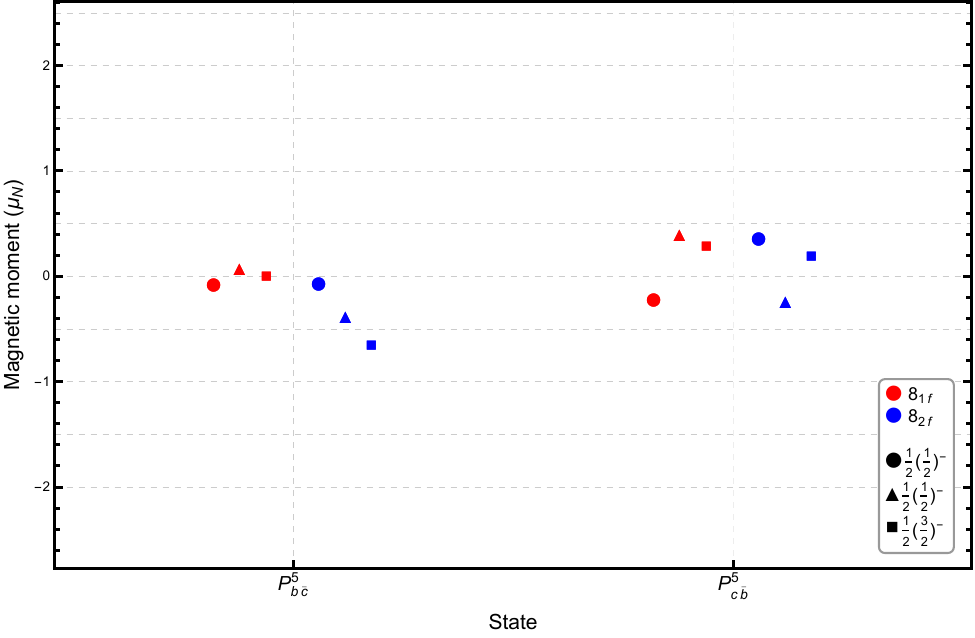}
\includegraphics[width=0.45\textwidth]{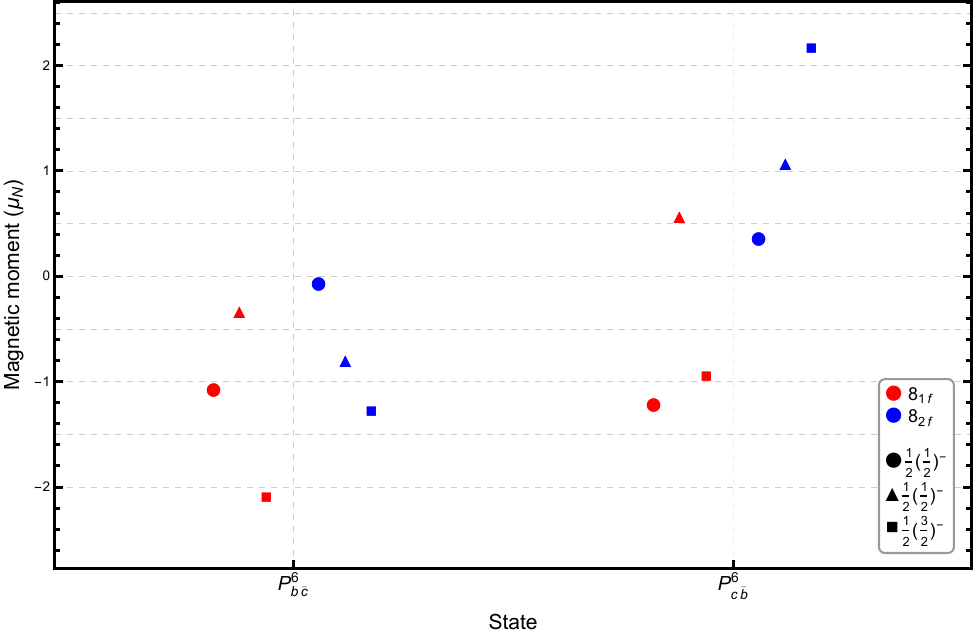}\\
\includegraphics[width=0.45\textwidth]{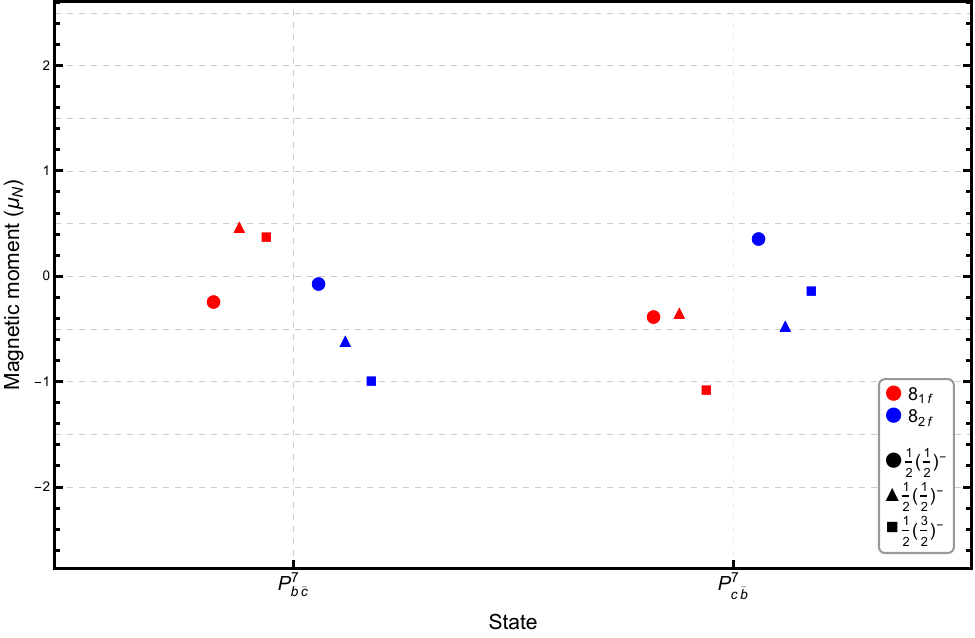}
\includegraphics[width=0.45\textwidth]{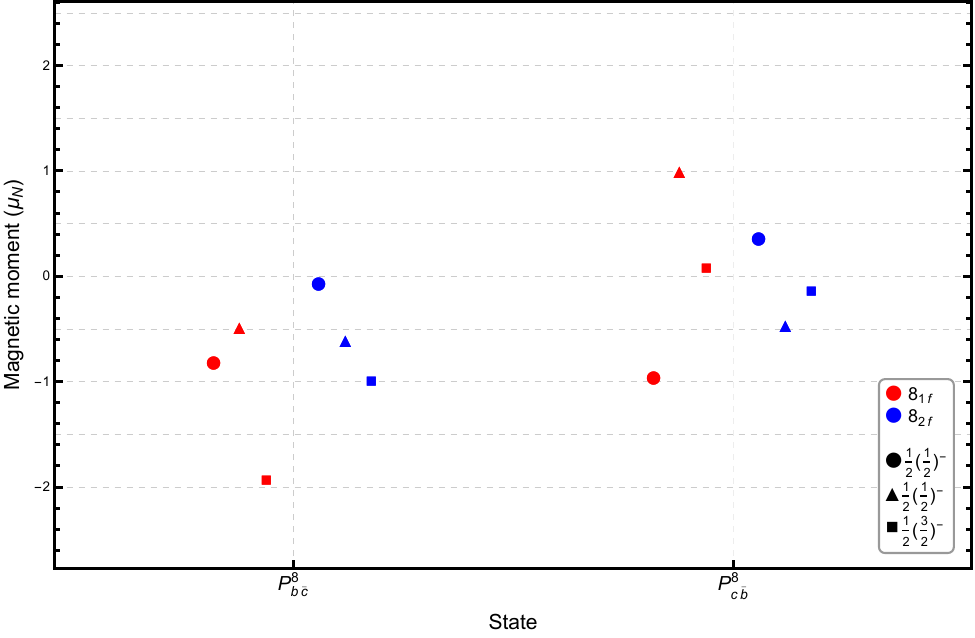}
\caption{Magnetic moments of the open bottom--charm molecular pentaquark octets $bc\bar{}$ and $c\bar{b}$ in the $8_{1f}$ and $8_{2f}$ flavor representations. The results are shown for the $J^{P}=\frac{1}{2}^{-}$ ($\frac{1}{2}^{+}\!\otimes\!0^{-}$ and $\frac{1}{2}^{+}\!\otimes\!1^{-}$)
and $J^{P}=\frac{3}{2}^{-}$ ($\frac{1}{2}^{+}\!\otimes\!1^{-}$) configurations,
with magnetic moments expressed in units of the nuclear magneton $\mu_N$. The red color refers to $8_{1f}$ representation while the blue color refers to the $8_{2f}$ representation. The spin-parity configurations in Eq.~\eqref{eq:JP} are denoted by circle, triangle, and square, respectively. }
 \label{fig:magmom}
  \end{figure}

Figure~\ref{fig:magmom} summarizes the magnetic moments in both representations and visually confirms the hierarchies discussed above: the near-universal values for the $8_{2f}$ pseudoscalar states, the broad dispersion in the $8_{1f}$ pseudoscalar states, the substantial shifts and sign changes upon introducing vector mesons, and the consistently largest magnitudes for the $J^{P}=\tfrac32^{-}$ configurations.

\section{Final Remarks}\label{final}

We have computed the magnetic moments of the open bottom--charm and charm--antibottom molecular pentaquark octets in the constituent quark model, using explicitly constructed spin--flavor wave functions of the two SU(3)$_f$ octet representations, $8_{1f}$ and $8_{2f}$. The results reveal a clear hierarchy. In the $8_{2f}$ representation, the $\tfrac12^+\!\otimes\!0^-$ states display near--universal values, $\mu = -0.062\,\mu_N$ for the entire $b\bar c$ octet and $\mu = +0.362\,\mu_N$ for the $c\bar b$ octet, reflecting the spin--singlet light--diquark structure that leaves the moment dominated by the heavy quark. The $8_{1f}$ representation, in contrast, exhibits a broad distribution of values with frequent sign changes. The inclusion of vector--meson components in the $\tfrac12^+\!\otimes\!1^-$ configurations introduces strong spin--dependent effects and sign reversals, and the $J^P=\tfrac32^-$ states systematically carry the largest magnitudes, making them the most favorable candidates for electromagnetic studies such as radiative transitions. The differences between the $b\bar c$ and $c\bar b$ families across all spin--parity assignments constitute direct evidence of heavy--quark flavor symmetry breaking in electromagnetic observables. Although direct measurements of static moments are challenging for short--lived pentaquarks, radiative widths and production asymmetries at LHCb, Belle~II, and the Electron--Ion Collider could provide indirect access, and the benchmarks presented here should help guide such searches.

\section*{Acknowledgment}
This work is supported by Scientific Research Projects Coordination Unit of Ondokuz Mayis University with project BAP05-2025-5384, and also supported in part by the National Natural Science Foundation of China under Project No. 12275023.

\bibliography{bcpenta}

\begin{thebibliography}{38}%
\makeatletter
\providecommand \@ifxundefined [1]{%
 \@ifx{#1\undefined}
}%
\providecommand \@ifnum [1]{%
 \ifnum #1\expandafter \@firstoftwo
 \else \expandafter \@secondoftwo
 \fi
}%
\providecommand \@ifx [1]{%
 \ifx #1\expandafter \@firstoftwo
 \else \expandafter \@secondoftwo
 \fi
}%
\providecommand \natexlab [1]{#1}%
\providecommand \enquote  [1]{``#1''}%
\providecommand \bibnamefont  [1]{#1}%
\providecommand \bibfnamefont [1]{#1}%
\providecommand \citenamefont [1]{#1}%
\providecommand \href@noop [0]{\@secondoftwo}%
\providecommand \href [0]{\begingroup \@sanitize@url \@href}%
\providecommand \@href[1]{\@@startlink{#1}\@@href}%
\providecommand \@@href[1]{\endgroup#1\@@endlink}%
\providecommand \@sanitize@url [0]{\catcode `\\12\catcode `\$12\catcode
  `\&12\catcode `\#12\catcode `\^12\catcode `\_12\catcode `\%12\relax}%
\providecommand \@@startlink[1]{}%
\providecommand \@@endlink[0]{}%
\providecommand \url  [0]{\begingroup\@sanitize@url \@url }%
\providecommand \@url [1]{\endgroup\@href {#1}{\urlprefix }}%
\providecommand \urlprefix  [0]{URL }%
\providecommand \Eprint [0]{\href }%
\providecommand \doibase [0]{https://doi.org/}%
\providecommand \selectlanguage [0]{\@gobble}%
\providecommand \bibinfo  [0]{\@secondoftwo}%
\providecommand \bibfield  [0]{\@secondoftwo}%
\providecommand \translation [1]{[#1]}%
\providecommand \BibitemOpen [0]{}%
\providecommand \bibitemStop [0]{}%
\providecommand \bibitemNoStop [0]{.\EOS\space}%
\providecommand \EOS [0]{\spacefactor3000\relax}%
\providecommand \BibitemShut  [1]{\csname bibitem#1\endcsname}%
\let\auto@bib@innerbib\@empty
\bibitem [{\citenamefont {Aaij}\ \emph {et~al.}(2015)\citenamefont {Aaij} \emph
  {et~al.}}]{LHCb:2015yax}%
  \BibitemOpen
  \bibfield  {author} {\bibinfo {author} {\bibfnamefont {R.}~\bibnamefont
  {Aaij}} \emph {et~al.} (\bibinfo {collaboration} {LHCb}),\ }\bibfield
  {title} {\bibinfo {title} {{Observation of $J/\psi p$ Resonances Consistent
  with Pentaquark States in $\Lambda_b^0 \to J/\psi K^- p$ Decays}},\ }\href
  {https://doi.org/10.1103/PhysRevLett.115.072001} {\bibfield  {journal}
  {\bibinfo  {journal} {Phys. Rev. Lett.}\ }\textbf {\bibinfo {volume} {115}},\
  \bibinfo {pages} {072001} (\bibinfo {year} {2015})},\ \Eprint
  {https://arxiv.org/abs/1507.03414} {arXiv:1507.03414 [hep-ex]} \BibitemShut
  {NoStop}%
\bibitem [{\citenamefont {Aaij}\ \emph {et~al.}(2019)\citenamefont {Aaij} \emph
  {et~al.}}]{LHCb:2019kea}%
  \BibitemOpen
  \bibfield  {author} {\bibinfo {author} {\bibfnamefont {R.}~\bibnamefont
  {Aaij}} \emph {et~al.} (\bibinfo {collaboration} {LHCb}),\ }\bibfield
  {title} {\bibinfo {title} {{Observation of a narrow pentaquark state,
  $P_c(4312)^+$, and of two-peak structure of the $P_c(4450)^+$}},\ }\href
  {https://doi.org/10.1103/PhysRevLett.122.222001} {\bibfield  {journal}
  {\bibinfo  {journal} {Phys. Rev. Lett.}\ }\textbf {\bibinfo {volume} {122}},\
  \bibinfo {pages} {222001} (\bibinfo {year} {2019})},\ \Eprint
  {https://arxiv.org/abs/1904.03947} {arXiv:1904.03947 [hep-ex]} \BibitemShut
  {NoStop}%
\bibitem [{\citenamefont {Gershon}(2022)}]{Gershon:2022xnn}%
  \BibitemOpen
  \bibfield  {author} {\bibinfo {author} {\bibfnamefont {T.}~\bibnamefont
  {Gershon}} (\bibinfo {collaboration} {LHCb}),\ }\bibfield  {title} {\bibinfo
  {title} {{Exotic hadron naming convention}}\ }\href
  {https://doi.org/10.17181/CERN.7XZO.HPH7} {10.17181/CERN.7XZO.HPH7} (\bibinfo
  {year} {2022}),\ \Eprint {https://arxiv.org/abs/2206.15233} {arXiv:2206.15233
  [hep-ex]} \BibitemShut {NoStop}%
\bibitem [{\citenamefont {Aaij}\ \emph {et~al.}(2021)\citenamefont {Aaij} \emph
  {et~al.}}]{LHCb:2020jpq}%
  \BibitemOpen
  \bibfield  {author} {\bibinfo {author} {\bibfnamefont {R.}~\bibnamefont
  {Aaij}} \emph {et~al.} (\bibinfo {collaboration} {LHCb}),\ }\bibfield
  {title} {\bibinfo {title} {{Evidence of a $J/\psi\Lambda$ structure and
  observation of excited $\Xi^-$ states in the $\Xi^-_b \to J/\psi\Lambda K^-$
  decay}},\ }\href {https://doi.org/10.1016/j.scib.2021.02.030} {\bibfield
  {journal} {\bibinfo  {journal} {Sci. Bull.}\ }\textbf {\bibinfo {volume}
  {66}},\ \bibinfo {pages} {1278} (\bibinfo {year} {2021})},\ \Eprint
  {https://arxiv.org/abs/2012.10380} {arXiv:2012.10380 [hep-ex]} \BibitemShut
  {NoStop}%
\bibitem [{\citenamefont {Adachi}\ \emph {et~al.}(2025)\citenamefont {Adachi}
  \emph {et~al.}}]{Belle:2025pey}%
  \BibitemOpen
  \bibfield  {author} {\bibinfo {author} {\bibfnamefont {I.}~\bibnamefont
  {Adachi}} \emph {et~al.} (\bibinfo {collaboration} {Belle, Belle-II}),\
  }\bibfield  {title} {\bibinfo {title} {{Search for Pcs(4459) and Pcs(4338) in
  Upsilon(1S,2S) inclusive decays at Belle}},\ }\href
  {https://doi.org/10.1103/pf8m-6j69} {\bibfield  {journal} {\bibinfo
  {journal} {Phys. Rev. Lett.}\ }\textbf {\bibinfo {volume} {135}},\ \bibinfo
  {pages} {041901} (\bibinfo {year} {2025})},\ \Eprint
  {https://arxiv.org/abs/2502.09951} {arXiv:2502.09951 [hep-ex]} \BibitemShut
  {NoStop}%
\bibitem [{\citenamefont {Aaij}\ \emph {et~al.}(2023)\citenamefont {Aaij} \emph
  {et~al.}}]{LHCb:2022ogu}%
  \BibitemOpen
  \bibfield  {author} {\bibinfo {author} {\bibfnamefont {R.}~\bibnamefont
  {Aaij}} \emph {et~al.} (\bibinfo {collaboration} {LHCb}),\ }\bibfield
  {title} {\bibinfo {title} {{Observation of a
  J/{\ensuremath{\psi}}{\ensuremath{\Lambda}} Resonance Consistent with a
  Strange Pentaquark Candidate in
  B-{\textrightarrow}J/{\ensuremath{\psi}}{\ensuremath{\Lambda}}p{\textasciimacron}
  Decays}},\ }\href {https://doi.org/10.1103/PhysRevLett.131.031901} {\bibfield
   {journal} {\bibinfo  {journal} {Phys. Rev. Lett.}\ }\textbf {\bibinfo
  {volume} {131}},\ \bibinfo {pages} {031901} (\bibinfo {year} {2023})},\
  \Eprint {https://arxiv.org/abs/2210.10346} {arXiv:2210.10346 [hep-ex]}
  \BibitemShut {NoStop}%
\bibitem [{\citenamefont {Chen}\ \emph {et~al.}(2017)\citenamefont {Chen},
  \citenamefont {He},\ and\ \citenamefont {Liu}}]{Chen:2016ryt}%
  \BibitemOpen
  \bibfield  {author} {\bibinfo {author} {\bibfnamefont {R.}~\bibnamefont
  {Chen}}, \bibinfo {author} {\bibfnamefont {J.}~\bibnamefont {He}},\ and\
  \bibinfo {author} {\bibfnamefont {X.}~\bibnamefont {Liu}},\ }\bibfield
  {title} {\bibinfo {title} {{Possible strange hidden-charm pentaquarks from
  $\Sigma_c^{(*)}\bar{D}_s^*$ and $\Xi^{(',*)}_c\bar{D}^*$ interactions}},\
  }\href {https://doi.org/10.1088/1674-1137/41/10/103105} {\bibfield  {journal}
  {\bibinfo  {journal} {Chin. Phys. C}\ }\textbf {\bibinfo {volume} {41}},\
  \bibinfo {pages} {103105} (\bibinfo {year} {2017})},\ \Eprint
  {https://arxiv.org/abs/1609.03235} {arXiv:1609.03235 [hep-ph]} \BibitemShut
  {NoStop}%
\bibitem [{\citenamefont {Wang}\ \emph {et~al.}(2020)\citenamefont {Wang},
  \citenamefont {Meng},\ and\ \citenamefont {Zhu}}]{Wang:2019nvm}%
  \BibitemOpen
  \bibfield  {author} {\bibinfo {author} {\bibfnamefont {B.}~\bibnamefont
  {Wang}}, \bibinfo {author} {\bibfnamefont {L.}~\bibnamefont {Meng}},\ and\
  \bibinfo {author} {\bibfnamefont {S.-L.}\ \bibnamefont {Zhu}},\ }\bibfield
  {title} {\bibinfo {title} {{Spectrum of the strange hidden charm molecular
  pentaquarks in chiral effective field theory}},\ }\href
  {https://doi.org/10.1103/PhysRevD.101.034018} {\bibfield  {journal} {\bibinfo
   {journal} {Phys. Rev. D}\ }\textbf {\bibinfo {volume} {101}},\ \bibinfo
  {pages} {034018} (\bibinfo {year} {2020})},\ \Eprint
  {https://arxiv.org/abs/1912.12592} {arXiv:1912.12592 [hep-ph]} \BibitemShut
  {NoStop}%
\bibitem [{\citenamefont {Yan}\ \emph {et~al.}(2023)\citenamefont {Yan},
  \citenamefont {Peng}, \citenamefont {S{\'a}nchez~S{\'a}nchez},\ and\
  \citenamefont {Pavon~Valderrama}}]{Yan:2022wuz}%
  \BibitemOpen
  \bibfield  {author} {\bibinfo {author} {\bibfnamefont {M.-J.}\ \bibnamefont
  {Yan}}, \bibinfo {author} {\bibfnamefont {F.-Z.}\ \bibnamefont {Peng}},
  \bibinfo {author} {\bibfnamefont {M.}~\bibnamefont
  {S{\'a}nchez~S{\'a}nchez}},\ and\ \bibinfo {author} {\bibfnamefont
  {M.}~\bibnamefont {Pavon~Valderrama}},\ }\bibfield  {title} {\bibinfo {title}
  {{P{\ensuremath{\psi}}s{\ensuremath{\Lambda}}(4338) pentaquark and its
  partners in the molecular picture}},\ }\href
  {https://doi.org/10.1103/PhysRevD.107.074025} {\bibfield  {journal} {\bibinfo
   {journal} {Phys. Rev. D}\ }\textbf {\bibinfo {volume} {107}},\ \bibinfo
  {pages} {074025} (\bibinfo {year} {2023})},\ \Eprint
  {https://arxiv.org/abs/2207.11144} {arXiv:2207.11144 [hep-ph]} \BibitemShut
  {NoStop}%
\bibitem [{\citenamefont {Wang}\ \emph {et~al.}(2016)\citenamefont {Wang},
  \citenamefont {Chen}, \citenamefont {Ma}, \citenamefont {Liu},\ and\
  \citenamefont {Zhu}}]{Wang:2016dzu}%
  \BibitemOpen
  \bibfield  {author} {\bibinfo {author} {\bibfnamefont {G.-J.}\ \bibnamefont
  {Wang}}, \bibinfo {author} {\bibfnamefont {R.}~\bibnamefont {Chen}}, \bibinfo
  {author} {\bibfnamefont {L.}~\bibnamefont {Ma}}, \bibinfo {author}
  {\bibfnamefont {X.}~\bibnamefont {Liu}},\ and\ \bibinfo {author}
  {\bibfnamefont {S.-L.}\ \bibnamefont {Zhu}},\ }\bibfield  {title} {\bibinfo
  {title} {{Magnetic moments of the hidden-charm pentaquark states}},\ }\href
  {https://doi.org/10.1103/PhysRevD.94.094018} {\bibfield  {journal} {\bibinfo
  {journal} {Phys. Rev. D}\ }\textbf {\bibinfo {volume} {94}},\ \bibinfo
  {pages} {094018} (\bibinfo {year} {2016})},\ \Eprint
  {https://arxiv.org/abs/1605.01337} {arXiv:1605.01337 [hep-ph]} \BibitemShut
  {NoStop}%
\bibitem [{\citenamefont {Ortiz-Pacheco}\ \emph {et~al.}(2019)\citenamefont
  {Ortiz-Pacheco}, \citenamefont {Bijker},\ and\ \citenamefont
  {Fern{\'a}ndez-Ram{\'\i}rez}}]{Ortiz-Pacheco:2018ccl}%
  \BibitemOpen
  \bibfield  {author} {\bibinfo {author} {\bibfnamefont {E.}~\bibnamefont
  {Ortiz-Pacheco}}, \bibinfo {author} {\bibfnamefont {R.}~\bibnamefont
  {Bijker}},\ and\ \bibinfo {author} {\bibfnamefont {C.}~\bibnamefont
  {Fern{\'a}ndez-Ram{\'\i}rez}},\ }\bibfield  {title} {\bibinfo {title}
  {{Hidden charm pentaquarks: mass spectrum, magnetic moments, and
  photocouplings}},\ }\href {https://doi.org/10.1088/1361-6471/ab096d}
  {\bibfield  {journal} {\bibinfo  {journal} {J. Phys. G}\ }\textbf {\bibinfo
  {volume} {46}},\ \bibinfo {pages} {065104} (\bibinfo {year} {2019})},\
  \Eprint {https://arxiv.org/abs/1808.10512} {arXiv:1808.10512 [nucl-th]}
  \BibitemShut {NoStop}%
\bibitem [{\citenamefont {Xu}\ \emph {et~al.}(2021)\citenamefont {Xu},
  \citenamefont {Liu},\ and\ \citenamefont {Huang}}]{Xu:2020flp}%
  \BibitemOpen
  \bibfield  {author} {\bibinfo {author} {\bibfnamefont {Y.-J.}\ \bibnamefont
  {Xu}}, \bibinfo {author} {\bibfnamefont {Y.-L.}\ \bibnamefont {Liu}},\ and\
  \bibinfo {author} {\bibfnamefont {M.-Q.}\ \bibnamefont {Huang}},\ }\bibfield
  {title} {\bibinfo {title} {{The magnetic moment of $P_{c}(4312)$ as a
  $\bar{D}\Sigma _{c}$ molecular state}},\ }\href
  {https://doi.org/10.1140/epjc/s10052-021-09211-8} {\bibfield  {journal}
  {\bibinfo  {journal} {Eur. Phys. J. C}\ }\textbf {\bibinfo {volume} {81}},\
  \bibinfo {pages} {421} (\bibinfo {year} {2021})},\ \Eprint
  {https://arxiv.org/abs/2008.07937} {arXiv:2008.07937 [hep-ph]} \BibitemShut
  {NoStop}%
\bibitem [{\citenamefont {{\"O}zdem}(2021)}]{Ozdem:2021ugy}%
  \BibitemOpen
  \bibfield  {author} {\bibinfo {author} {\bibfnamefont {U.}~\bibnamefont
  {{\"O}zdem}},\ }\bibfield  {title} {\bibinfo {title} {{Magnetic dipole
  moments of the hidden-charm pentaquark states: $P_c(4440)$, $P_c(4457)$ and
  $P_{cs}(4459)$}},\ }\href {https://doi.org/10.1140/epjc/s10052-021-09070-3}
  {\bibfield  {journal} {\bibinfo  {journal} {Eur. Phys. J. C}\ }\textbf
  {\bibinfo {volume} {81}},\ \bibinfo {pages} {277} (\bibinfo {year} {2021})},\
  \Eprint {https://arxiv.org/abs/2102.01996} {arXiv:2102.01996 [hep-ph]}
  \BibitemShut {NoStop}%
\bibitem [{\citenamefont {Li}\ \emph {et~al.}(2021)\citenamefont {Li},
  \citenamefont {Liu}, \citenamefont {Sun},\ and\ \citenamefont
  {Chen}}]{Li:2021ryu}%
  \BibitemOpen
  \bibfield  {author} {\bibinfo {author} {\bibfnamefont {M.-W.}\ \bibnamefont
  {Li}}, \bibinfo {author} {\bibfnamefont {Z.-W.}\ \bibnamefont {Liu}},
  \bibinfo {author} {\bibfnamefont {Z.-F.}\ \bibnamefont {Sun}},\ and\ \bibinfo
  {author} {\bibfnamefont {R.}~\bibnamefont {Chen}},\ }\bibfield  {title}
  {\bibinfo {title} {{Magnetic moments and transition magnetic moments of Pc
  and Pcs states}},\ }\href {https://doi.org/10.1103/PhysRevD.104.054016}
  {\bibfield  {journal} {\bibinfo  {journal} {Phys. Rev. D}\ }\textbf {\bibinfo
  {volume} {104}},\ \bibinfo {pages} {054016} (\bibinfo {year} {2021})},\
  \Eprint {https://arxiv.org/abs/2106.15053} {arXiv:2106.15053 [hep-ph]}
  \BibitemShut {NoStop}%
\bibitem [{\citenamefont {{\"O}zdem}(2023{\natexlab{a}})}]{Ozdem:2023htj}%
  \BibitemOpen
  \bibfield  {author} {\bibinfo {author} {\bibfnamefont {U.}~\bibnamefont
  {{\"O}zdem}},\ }\bibfield  {title} {\bibinfo {title} {{Electromagnetic
  properties of
  D{\textasciimacron}({\textasteriskcentered}){\ensuremath{\Xi}}c',
  D{\textasciimacron}({\textasteriskcentered}){\ensuremath{\Lambda}}c,
  D{\textasciimacron}s({\textasteriskcentered}){\ensuremath{\Lambda}}c and
  D{\textasciimacron}s({\textasteriskcentered}){\ensuremath{\Xi}}c
  pentaquarks}},\ }\href {https://doi.org/10.1016/j.physletb.2023.138267}
  {\bibfield  {journal} {\bibinfo  {journal} {Phys. Lett. B}\ }\textbf
  {\bibinfo {volume} {846}},\ \bibinfo {pages} {138267} (\bibinfo {year}
  {2023}{\natexlab{a}})},\ \Eprint {https://arxiv.org/abs/2303.10649}
  {arXiv:2303.10649 [hep-ph]} \BibitemShut {NoStop}%
\bibitem [{\citenamefont {{\"O}zdem}(2023{\natexlab{b}})}]{Ozdem:2022kei}%
  \BibitemOpen
  \bibfield  {author} {\bibinfo {author} {\bibfnamefont {U.}~\bibnamefont
  {{\"O}zdem}},\ }\bibfield  {title} {\bibinfo {title} {{Investigation of
  magnetic moment of Pcs(4338) and Pcs(4459) pentaquark states}},\ }\href
  {https://doi.org/10.1016/j.physletb.2022.137635} {\bibfield  {journal}
  {\bibinfo  {journal} {Phys. Lett. B}\ }\textbf {\bibinfo {volume} {836}},\
  \bibinfo {pages} {137635} (\bibinfo {year} {2023}{\natexlab{b}})},\ \Eprint
  {https://arxiv.org/abs/2208.07684} {arXiv:2208.07684 [hep-ph]} \BibitemShut
  {NoStop}%
\bibitem [{\citenamefont {Gao}\ and\ \citenamefont {Li}(2022)}]{Gao:2021hmv}%
  \BibitemOpen
  \bibfield  {author} {\bibinfo {author} {\bibfnamefont {F.}~\bibnamefont
  {Gao}}\ and\ \bibinfo {author} {\bibfnamefont {H.-S.}\ \bibnamefont {Li}},\
  }\bibfield  {title} {\bibinfo {title} {{Magnetic moments of hidden-charm
  strange pentaquark states*}},\ }\href
  {https://doi.org/10.1088/1674-1137/ac8651} {\bibfield  {journal} {\bibinfo
  {journal} {Chin. Phys. C}\ }\textbf {\bibinfo {volume} {46}},\ \bibinfo
  {pages} {123111} (\bibinfo {year} {2022})},\ \Eprint
  {https://arxiv.org/abs/2112.01823} {arXiv:2112.01823 [hep-ph]} \BibitemShut
  {NoStop}%
\bibitem [{\citenamefont {Guo}\ and\ \citenamefont {Li}(2024)}]{Guo:2023fih}%
  \BibitemOpen
  \bibfield  {author} {\bibinfo {author} {\bibfnamefont {F.}~\bibnamefont
  {Guo}}\ and\ \bibinfo {author} {\bibfnamefont {H.-S.}\ \bibnamefont {Li}},\
  }\bibfield  {title} {\bibinfo {title} {{Analysis of the hidden-charm
  pentaquark states based on magnetic moment and transition magnetic moment}},\
  }\href {https://doi.org/10.1140/epjc/s10052-024-12699-5} {\bibfield
  {journal} {\bibinfo  {journal} {Eur. Phys. J. C}\ }\textbf {\bibinfo {volume}
  {84}},\ \bibinfo {pages} {392} (\bibinfo {year} {2024})},\ \Eprint
  {https://arxiv.org/abs/2304.10981} {arXiv:2304.10981 [hep-ph]} \BibitemShut
  {NoStop}%
\bibitem [{\citenamefont {Wang}\ \emph {et~al.}(2023)\citenamefont {Wang},
  \citenamefont {Luo}, \citenamefont {Zhou}, \citenamefont {Liu},\ and\
  \citenamefont {Liu}}]{Wang:2022nqs}%
  \BibitemOpen
  \bibfield  {author} {\bibinfo {author} {\bibfnamefont {F.-L.}\ \bibnamefont
  {Wang}}, \bibinfo {author} {\bibfnamefont {S.-Q.}\ \bibnamefont {Luo}},
  \bibinfo {author} {\bibfnamefont {H.-Y.}\ \bibnamefont {Zhou}}, \bibinfo
  {author} {\bibfnamefont {Z.-W.}\ \bibnamefont {Liu}},\ and\ \bibinfo {author}
  {\bibfnamefont {X.}~\bibnamefont {Liu}},\ }\bibfield  {title} {\bibinfo
  {title} {{Exploring the electromagnetic properties of the
  {\ensuremath{\Xi}}c(',*)D{\textasciimacron}s* and
  {\ensuremath{\Omega}}c(*)D{\textasciimacron}s* molecular states}},\ }\href
  {https://doi.org/10.1103/PhysRevD.108.034006} {\bibfield  {journal} {\bibinfo
   {journal} {Phys. Rev. D}\ }\textbf {\bibinfo {volume} {108}},\ \bibinfo
  {pages} {034006} (\bibinfo {year} {2023})},\ \Eprint
  {https://arxiv.org/abs/2210.02809} {arXiv:2210.02809 [hep-ph]} \BibitemShut
  {NoStop}%
\bibitem [{\citenamefont {Wang}\ \emph {et~al.}(2022)\citenamefont {Wang},
  \citenamefont {Zhou}, \citenamefont {Liu},\ and\ \citenamefont
  {Liu}}]{Wang:2022tib}%
  \BibitemOpen
  \bibfield  {author} {\bibinfo {author} {\bibfnamefont {F.-L.}\ \bibnamefont
  {Wang}}, \bibinfo {author} {\bibfnamefont {H.-Y.}\ \bibnamefont {Zhou}},
  \bibinfo {author} {\bibfnamefont {Z.-W.}\ \bibnamefont {Liu}},\ and\ \bibinfo
  {author} {\bibfnamefont {X.}~\bibnamefont {Liu}},\ }\bibfield  {title}
  {\bibinfo {title} {{What can we learn from the electromagnetic properties of
  hidden-charm molecular pentaquarks with single strangeness?}},\ }\href
  {https://doi.org/10.1103/PhysRevD.106.054020} {\bibfield  {journal} {\bibinfo
   {journal} {Phys. Rev. D}\ }\textbf {\bibinfo {volume} {106}},\ \bibinfo
  {pages} {054020} (\bibinfo {year} {2022})},\ \Eprint
  {https://arxiv.org/abs/2208.10756} {arXiv:2208.10756 [hep-ph]} \BibitemShut
  {NoStop}%
\bibitem [{\citenamefont {{\"O}zdem}(2024)}]{Ozdem:2024jty}%
  \BibitemOpen
  \bibfield  {author} {\bibinfo {author} {\bibfnamefont {U.}~\bibnamefont
  {{\"O}zdem}},\ }\bibfield  {title} {\bibinfo {title} {{Analysis of the
  isospin eigenstate $\bar{D} \Sigma _c$, $\bar{D}^{*} \Sigma _c$, and $\bar{D}
  \Sigma _c^{*}$ pentaquarks by their electromagnetic properties}},\ }\href
  {https://doi.org/10.1140/epjc/s10052-024-13124-7} {\bibfield  {journal}
  {\bibinfo  {journal} {Eur. Phys. J. C}\ }\textbf {\bibinfo {volume} {84}},\
  \bibinfo {pages} {769} (\bibinfo {year} {2024})},\ \Eprint
  {https://arxiv.org/abs/2401.12678} {arXiv:2401.12678 [hep-ph]} \BibitemShut
  {NoStop}%
\bibitem [{\citenamefont {Li}\ \emph {et~al.}(2024)\citenamefont {Li},
  \citenamefont {Guo}, \citenamefont {Lei},\ and\ \citenamefont
  {Gao}}]{Li:2024wxr}%
  \BibitemOpen
  \bibfield  {author} {\bibinfo {author} {\bibfnamefont {H.-S.}\ \bibnamefont
  {Li}}, \bibinfo {author} {\bibfnamefont {F.}~\bibnamefont {Guo}}, \bibinfo
  {author} {\bibfnamefont {Y.-D.}\ \bibnamefont {Lei}},\ and\ \bibinfo {author}
  {\bibfnamefont {F.}~\bibnamefont {Gao}},\ }\bibfield  {title} {\bibinfo
  {title} {{Magnetic moments and axial charges of the octet hidden-charm
  molecular pentaquark family}},\ }\href
  {https://doi.org/10.1103/PhysRevD.109.094027} {\bibfield  {journal} {\bibinfo
   {journal} {Phys. Rev. D}\ }\textbf {\bibinfo {volume} {109}},\ \bibinfo
  {pages} {094027} (\bibinfo {year} {2024})},\ \Eprint
  {https://arxiv.org/abs/2401.14767} {arXiv:2401.14767 [hep-ph]} \BibitemShut
  {NoStop}%
\bibitem [{\citenamefont {Li}(2024)}]{Li:2024jlq}%
  \BibitemOpen
  \bibfield  {author} {\bibinfo {author} {\bibfnamefont {H.-S.}\ \bibnamefont
  {Li}},\ }\bibfield  {title} {\bibinfo {title} {{Molecular pentaquark magnetic
  moments in heavy pentaquark chiral perturbation theory}},\ }\href
  {https://doi.org/10.1103/PhysRevD.109.114039} {\bibfield  {journal} {\bibinfo
   {journal} {Phys. Rev. D}\ }\textbf {\bibinfo {volume} {109}},\ \bibinfo
  {pages} {114039} (\bibinfo {year} {2024})},\ \Eprint
  {https://arxiv.org/abs/2401.14759} {arXiv:2401.14759 [hep-ph]} \BibitemShut
  {NoStop}%
\bibitem [{\citenamefont {{\"O}zdem}(2025)}]{Ozdem:2025fks}%
  \BibitemOpen
  \bibfield  {author} {\bibinfo {author} {\bibfnamefont {U.}~\bibnamefont
  {{\"O}zdem}},\ }\bibfield  {title} {\bibinfo {title} {{Probing the
  electromagnetic structure of the $P_c(4337)^+$ pentaquark: insights from a
  diquark{\textendash}diquark{\textendash}antiquark picture for $J^P =
  \frac{1}{2}^-$ and $\frac{3}{2}^-$ states}},\ }\href
  {https://doi.org/10.1140/epjc/s10052-025-14439-9} {\bibfield  {journal}
  {\bibinfo  {journal} {Eur. Phys. J. C}\ }\textbf {\bibinfo {volume} {85}},\
  \bibinfo {pages} {704} (\bibinfo {year} {2025})},\ \Eprint
  {https://arxiv.org/abs/2506.04345} {arXiv:2506.04345 [hep-ph]} \BibitemShut
  {NoStop}%
\bibitem [{\citenamefont {{\"O}zdem}(2026)}]{Ozdem:2025jda}%
  \BibitemOpen
  \bibfield  {author} {\bibinfo {author} {\bibfnamefont {U.}~\bibnamefont
  {{\"O}zdem}},\ }\bibfield  {title} {\bibinfo {title} {{Electromagnetic
  tomography of spin-$ \frac{3}{2} $ hidden-charm strange pentaquarks}},\
  }\href {https://doi.org/10.1007/JHEP02(2026)207} {\bibfield  {journal}
  {\bibinfo  {journal} {JHEP}\ }\textbf {\bibinfo {volume} {02}},\ \bibinfo
  {pages} {207}},\ \Eprint {https://arxiv.org/abs/2510.26893} {arXiv:2510.26893
  [hep-ph]} \BibitemShut {NoStop}%
\bibitem [{\citenamefont {Mutuk}(2024)}]{Mutuk:2024jxf}%
  \BibitemOpen
  \bibfield  {author} {\bibinfo {author} {\bibfnamefont {H.}~\bibnamefont
  {Mutuk}},\ }\bibfield  {title} {\bibinfo {title} {{Magnetic moments of
  hidden-bottom pentaquark states}},\ }\href
  {https://doi.org/10.1140/epjc/s10052-024-13263-x} {\bibfield  {journal}
  {\bibinfo  {journal} {Eur. Phys. J. C}\ }\textbf {\bibinfo {volume} {84}},\
  \bibinfo {pages} {874} (\bibinfo {year} {2024})},\ \Eprint
  {https://arxiv.org/abs/2403.16616} {arXiv:2403.16616 [hep-ph]} \BibitemShut
  {NoStop}%
\bibitem [{\citenamefont {Mutuk}\ and\ \citenamefont
  {Kang}(2024)}]{Mutuk:2024ltc}%
  \BibitemOpen
  \bibfield  {author} {\bibinfo {author} {\bibfnamefont {H.}~\bibnamefont
  {Mutuk}}\ and\ \bibinfo {author} {\bibfnamefont {X.-W.}\ \bibnamefont
  {Kang}},\ }\bibfield  {title} {\bibinfo {title} {{Unveiling the structure of
  hidden-bottom strange pentaquarks via magnetic moments}},\ }\href
  {https://doi.org/10.1016/j.physletb.2024.138772} {\bibfield  {journal}
  {\bibinfo  {journal} {Phys. Lett. B}\ }\textbf {\bibinfo {volume} {855}},\
  \bibinfo {pages} {138772} (\bibinfo {year} {2024})},\ \Eprint
  {https://arxiv.org/abs/2405.07066} {arXiv:2405.07066 [hep-ph]} \BibitemShut
  {NoStop}%
\bibitem [{\citenamefont {Wu}\ \emph {et~al.}(2017)\citenamefont {Wu},
  \citenamefont {Liu}, \citenamefont {Chen}, \citenamefont {Liu},\ and\
  \citenamefont {Zhu}}]{Wu:2017weo}%
  \BibitemOpen
  \bibfield  {author} {\bibinfo {author} {\bibfnamefont {J.}~\bibnamefont
  {Wu}}, \bibinfo {author} {\bibfnamefont {Y.-R.}\ \bibnamefont {Liu}},
  \bibinfo {author} {\bibfnamefont {K.}~\bibnamefont {Chen}}, \bibinfo {author}
  {\bibfnamefont {X.}~\bibnamefont {Liu}},\ and\ \bibinfo {author}
  {\bibfnamefont {S.-L.}\ \bibnamefont {Zhu}},\ }\bibfield  {title} {\bibinfo
  {title} {{Hidden-charm pentaquarks and their hidden-bottom and $B_c$-like
  partner states}},\ }\href {https://doi.org/10.1103/PhysRevD.95.034002}
  {\bibfield  {journal} {\bibinfo  {journal} {Phys. Rev. D}\ }\textbf {\bibinfo
  {volume} {95}},\ \bibinfo {pages} {034002} (\bibinfo {year} {2017})},\
  \Eprint {https://arxiv.org/abs/1701.03873} {arXiv:1701.03873 [hep-ph]}
  \BibitemShut {NoStop}%
\bibitem [{\citenamefont {Lin}\ \emph {et~al.}(2024)\citenamefont {Lin},
  \citenamefont {Chen}, \citenamefont {Liang}, \citenamefont {Liu},\ and\
  \citenamefont {Zhou}}]{Lin:2023iww}%
  \BibitemOpen
  \bibfield  {author} {\bibinfo {author} {\bibfnamefont {J.-X.}\ \bibnamefont
  {Lin}}, \bibinfo {author} {\bibfnamefont {H.-X.}\ \bibnamefont {Chen}},
  \bibinfo {author} {\bibfnamefont {W.-H.}\ \bibnamefont {Liang}}, \bibinfo
  {author} {\bibfnamefont {W.-Y.}\ \bibnamefont {Liu}},\ and\ \bibinfo {author}
  {\bibfnamefont {D.}~\bibnamefont {Zhou}},\ }\bibfield  {title} {\bibinfo
  {title} {{Molecular pentaquark states with open charm and bottom flavors}},\
  }\href {https://doi.org/10.1140/epja/s10050-024-01240-7} {\bibfield
  {journal} {\bibinfo  {journal} {Eur. Phys. J. A}\ }\textbf {\bibinfo {volume}
  {60}},\ \bibinfo {pages} {15} (\bibinfo {year} {2024})},\ \Eprint
  {https://arxiv.org/abs/2308.01007} {arXiv:2308.01007 [hep-ph]} \BibitemShut
  {NoStop}%
\bibitem [{\citenamefont {Wang}\ and\ \citenamefont
  {Long}(2025)}]{Wang:2025hhx}%
  \BibitemOpen
  \bibfield  {author} {\bibinfo {author} {\bibfnamefont {Z.-Y.}\ \bibnamefont
  {Wang}}\ and\ \bibinfo {author} {\bibfnamefont {Z.-W.}\ \bibnamefont
  {Long}},\ }\bibfield  {title} {\bibinfo {title} {{Prediction of
  QQqqs{\textasciimacron} molecular pentaquarks within the extended local
  hidden gauge approach}},\ }\href {https://doi.org/10.1103/zf47-xh7z}
  {\bibfield  {journal} {\bibinfo  {journal} {Phys. Rev. D}\ }\textbf {\bibinfo
  {volume} {112}},\ \bibinfo {pages} {056029} (\bibinfo {year} {2025})},\
  \Eprint {https://arxiv.org/abs/2508.21474} {arXiv:2508.21474 [hep-ph]}
  \BibitemShut {NoStop}%
\bibitem [{\citenamefont {Navas}\ \emph {et~al.}(2024)\citenamefont {Navas}
  \emph {et~al.}}]{ParticleDataGroup:2024cfk}%
  \BibitemOpen
  \bibfield  {author} {\bibinfo {author} {\bibfnamefont {S.}~\bibnamefont
  {Navas}} \emph {et~al.} (\bibinfo {collaboration} {Particle Data Group}),\
  }\bibfield  {title} {\bibinfo {title} {{Review of particle physics}},\ }\href
  {https://doi.org/10.1103/PhysRevD.110.030001} {\bibfield  {journal} {\bibinfo
   {journal} {Phys. Rev. D}\ }\textbf {\bibinfo {volume} {110}},\ \bibinfo
  {pages} {030001} (\bibinfo {year} {2024})}\BibitemShut {NoStop}%
\bibitem [{\citenamefont {Yang}\ \emph {et~al.}(2012)\citenamefont {Yang},
  \citenamefont {Sun}, \citenamefont {He}, \citenamefont {Liu},\ and\
  \citenamefont {Zhu}}]{Yang:2011wz}%
  \BibitemOpen
  \bibfield  {author} {\bibinfo {author} {\bibfnamefont {Z.-C.}\ \bibnamefont
  {Yang}}, \bibinfo {author} {\bibfnamefont {Z.-F.}\ \bibnamefont {Sun}},
  \bibinfo {author} {\bibfnamefont {J.}~\bibnamefont {He}}, \bibinfo {author}
  {\bibfnamefont {X.}~\bibnamefont {Liu}},\ and\ \bibinfo {author}
  {\bibfnamefont {S.-L.}\ \bibnamefont {Zhu}},\ }\bibfield  {title} {\bibinfo
  {title} {{The possible hidden-charm molecular baryons composed of
  anti-charmed meson and charmed baryon}},\ }\href
  {https://doi.org/10.1088/1674-1137/36/1/002} {\bibfield  {journal} {\bibinfo
  {journal} {Chin. Phys. C}\ }\textbf {\bibinfo {volume} {36}},\ \bibinfo
  {pages} {6} (\bibinfo {year} {2012})},\ \Eprint
  {https://arxiv.org/abs/1105.2901} {arXiv:1105.2901 [hep-ph]} \BibitemShut
  {NoStop}%
\bibitem [{\citenamefont {Jacob}\ and\ \citenamefont
  {Wick}(1959)}]{Jacob:1959at}%
  \BibitemOpen
  \bibfield  {author} {\bibinfo {author} {\bibfnamefont {M.}~\bibnamefont
  {Jacob}}\ and\ \bibinfo {author} {\bibfnamefont {G.~C.}\ \bibnamefont
  {Wick}},\ }\bibfield  {title} {\bibinfo {title} {{On the General Theory of
  Collisions for Particles with Spin}},\ }\href
  {https://doi.org/10.1006/aphy.2000.6022} {\bibfield  {journal} {\bibinfo
  {journal} {Annals Phys.}\ }\textbf {\bibinfo {volume} {7}},\ \bibinfo {pages}
  {404} (\bibinfo {year} {1959})}\BibitemShut {NoStop}%
\bibitem [{\citenamefont {Aaij}\ \emph {et~al.}(2025)\citenamefont {Aaij} \emph
  {et~al.}}]{LHCb:2024eyx}%
  \BibitemOpen
  \bibfield  {author} {\bibinfo {author} {\bibfnamefont {R.}~\bibnamefont
  {Aaij}} \emph {et~al.} (\bibinfo {collaboration} {LHCb}),\ }\bibfield
  {title} {\bibinfo {title} {{First Determination of the Spin-Parity of
  {\ensuremath{\Xi}}c(3055)+,0 Baryons}},\ }\href
  {https://doi.org/10.1103/PhysRevLett.134.081901} {\bibfield  {journal}
  {\bibinfo  {journal} {Phys. Rev. Lett.}\ }\textbf {\bibinfo {volume} {134}},\
  \bibinfo {pages} {081901} (\bibinfo {year} {2025})},\ \Eprint
  {https://arxiv.org/abs/2409.05440} {arXiv:2409.05440 [hep-ex]} \BibitemShut
  {NoStop}%
\bibitem [{\citenamefont {Kang}\ \emph {et~al.}(2010)\citenamefont {Kang},
  \citenamefont {Li},\ and\ \citenamefont {Zou}}]{Kang:2009sb}%
  \BibitemOpen
  \bibfield  {author} {\bibinfo {author} {\bibfnamefont {X.-W.}\ \bibnamefont
  {Kang}}, \bibinfo {author} {\bibfnamefont {H.-B.}\ \bibnamefont {Li}},\ and\
  \bibinfo {author} {\bibfnamefont {B.-S.}\ \bibnamefont {Zou}},\ }\bibfield
  {title} {\bibinfo {title} {{Partial wave analysis of psi' ---{\ensuremath{>}}
  gamma chi(c0) ---{\ensuremath{>}} gamma p K- Lambda-bar being used for
  searching for baryon resonance}},\ }\href
  {https://doi.org/10.1088/1674-1137/34/8/005} {\bibfield  {journal} {\bibinfo
  {journal} {Chin. Phys. C}\ }\textbf {\bibinfo {volume} {34}},\ \bibinfo
  {pages} {1061} (\bibinfo {year} {2010})},\ \Eprint
  {https://arxiv.org/abs/0911.2998} {arXiv:0911.2998 [hep-ph]} \BibitemShut
  {NoStop}%
\bibitem [{\citenamefont {Kang}\ \emph {et~al.}(2011)\citenamefont {Kang},
  \citenamefont {Li}, \citenamefont {Lu},\ and\ \citenamefont
  {Datta}}]{Kang:2010td}%
  \BibitemOpen
  \bibfield  {author} {\bibinfo {author} {\bibfnamefont {X.-W.}\ \bibnamefont
  {Kang}}, \bibinfo {author} {\bibfnamefont {H.-B.}\ \bibnamefont {Li}},
  \bibinfo {author} {\bibfnamefont {G.-R.}\ \bibnamefont {Lu}},\ and\ \bibinfo
  {author} {\bibfnamefont {A.}~\bibnamefont {Datta}},\ }\bibfield  {title}
  {\bibinfo {title} {{Study of CP violation in $\Lambda_c^+$ decay}},\ }\href
  {https://doi.org/10.1142/S0217751X11053432} {\bibfield  {journal} {\bibinfo
  {journal} {Int. J. Mod. Phys. A}\ }\textbf {\bibinfo {volume} {26}},\
  \bibinfo {pages} {2523} (\bibinfo {year} {2011})},\ \Eprint
  {https://arxiv.org/abs/1003.5494} {arXiv:1003.5494 [hep-ph]} \BibitemShut
  {NoStop}%
\bibitem [{\citenamefont {Charles}\ \emph {et~al.}(2010)\citenamefont
  {Charles}, \citenamefont {Descotes-Genon}, \citenamefont {Kang},
  \citenamefont {Li},\ and\ \citenamefont {Lu}}]{Charles:2009ig}%
  \BibitemOpen
  \bibfield  {author} {\bibinfo {author} {\bibfnamefont {J.}~\bibnamefont
  {Charles}}, \bibinfo {author} {\bibfnamefont {S.}~\bibnamefont
  {Descotes-Genon}}, \bibinfo {author} {\bibfnamefont {X.-W.}\ \bibnamefont
  {Kang}}, \bibinfo {author} {\bibfnamefont {H.-B.}\ \bibnamefont {Li}},\ and\
  \bibinfo {author} {\bibfnamefont {G.-R.}\ \bibnamefont {Lu}},\ }\bibfield
  {title} {\bibinfo {title} {{Extracting CP violation and strong phase in D
  decays by using quantum correlations in psi(3770) ---{\ensuremath{>}} D0
  D0-bar ---{\ensuremath{>}} (V(1)V(2))(V(3)V(4)) and psi(3770)
  ---{\ensuremath{>}} D0 D0-bar ----{\ensuremath{>}} (V(1)V(2))(K pi)}},\
  }\href {https://doi.org/10.1103/PhysRevD.81.054032} {\bibfield  {journal}
  {\bibinfo  {journal} {Phys. Rev. D}\ }\textbf {\bibinfo {volume} {81}},\
  \bibinfo {pages} {054032} (\bibinfo {year} {2010})},\ \Eprint
  {https://arxiv.org/abs/0912.0899} {arXiv:0912.0899 [hep-ph]} \BibitemShut
  {NoStop}%
\bibitem [{\citenamefont {Wang}\ \emph {et~al.}(2019)\citenamefont {Wang},
  \citenamefont {Yang}, \citenamefont {Meng},\ and\ \citenamefont
  {Zhu}}]{Wang:2019mhm}%
  \BibitemOpen
  \bibfield  {author} {\bibinfo {author} {\bibfnamefont {B.}~\bibnamefont
  {Wang}}, \bibinfo {author} {\bibfnamefont {B.}~\bibnamefont {Yang}}, \bibinfo
  {author} {\bibfnamefont {L.}~\bibnamefont {Meng}},\ and\ \bibinfo {author}
  {\bibfnamefont {S.-L.}\ \bibnamefont {Zhu}},\ }\bibfield  {title} {\bibinfo
  {title} {{Radiative transitions and magnetic moments of the charmed and
  bottom vector mesons in chiral perturbation theory}},\ }\href
  {https://doi.org/10.1103/PhysRevD.100.016019} {\bibfield  {journal} {\bibinfo
   {journal} {Phys. Rev. D}\ }\textbf {\bibinfo {volume} {100}},\ \bibinfo
  {pages} {016019} (\bibinfo {year} {2019})},\ \Eprint
  {https://arxiv.org/abs/1905.07742} {arXiv:1905.07742 [hep-ph]} \BibitemShut
  {NoStop}%
\end{thebibliography}%

\end{document}